\documentclass[twocolumn]{aastex631}

\usepackage[T1]{fontenc}

\shorttitle{}
\shortauthors{Hasegawa et al.}
\graphicspath{{./}{figures/}}

\begin{document}

\title{Testing magnetospheric accretion as an H$\alpha$ emission mechanism of embedded giant planets: \\
The case study for the disk exhibiting meridional flow around HD 163296}

\author[0000-0002-9017-3663]{Yasuhiro Hasegawa}
\affiliation{Jet Propulsion Laboratory, California Institute of Technology, Pasadena, CA 91109, USA}
\email{yasuhiro.hasegawa@jpl.nasa.gov}

\author[0000-0002-6879-3030]{Taichi Uyama}
    \affiliation{Department of Physics and Astronomy, California State University Northridge, 18111 Nordhoff Street, Northridge, CA 91330, USA} 
    \affiliation{Infrared Processing and Analysis Center, California Institute of Technology, 1200 E. California Blvd., Pasadena, CA 91125, USA}
    \affiliation{NASA Exoplanet Science Institute, Pasadena, CA 91125, USA}
    \affiliation{National Astronomical Observatory of Japan, 2-21-1 Osawa, Mitaka, Tokyo 181-8588, Japan}

\author[0000-0002-3053-3575]{Jun Hashimoto}
\affiliation{Astrobiology Center, National Institutes of Natural Sciences, 2-21-1 Osawa, Mitaka, Tokyo 181-8588, Japan}
\affiliation{Subaru Telescope, National Astronomical Observatory of Japan, Mitaka, Tokyo 181-8588, Japan}
\affiliation{Department of Astronomy, School of Science, Graduate University for Advanced Studies (SOKENDAI), Mitaka, Tokyo 181-8588, Japan}

\author[0000-0003-0568-9225]{Yuhiko Aoyama}
\affiliation{Kavli Institute for Astronomy and Astrophysics, Peking University, Yiheyuan Road 5, Haidian District, Beijing 100871, People’s Republic of China}

\author[0000-0003-4514-7906]{Vincent Deo}
\affiliation{Subaru Telescope, National Astronomical Observatory of Japan, National Institutes of Natural Sciences, 650 North A`oh$\bar{o}$k$\bar{u}$ Place, Hilo, HI 96720, USA}

\author[0000-0002-1097-9908]{Olivier Guyon}
\affiliation{Subaru Telescope, National Astronomical Observatory of Japan, National Institutes of Natural Sciences, 650 North A`oh$\bar{o}$k$\bar{u}$ Place, Hilo, HI 96720, USA}
\affiliation{Steward Observatory, University of Arizona, Tucson, AZ 85721, USA}
\affiliation{Astrobiology Center of NINS, 2-21-1 Osawa, Mitaka, Tokyo 181-8588, Japan}

\author[0000-0002-3047-1845]{Julien Lozi}
\affiliation{Subaru Telescope, National Astronomical Observatory of Japan, National Institutes of Natural Sciences, 650 North A`oh$\bar{o}$k$\bar{u}$ Place, Hilo, HI 96720, USA}
    
\author[0000-0002-8352-7515]{Barnaby Norris}
\affiliation{Sydney Institute for Astronomy, School of Physics, Physics Road, University of Sydney, NSW 2006, Australia}
 \affiliation{AAO-USyd, School of Physics, University of Sydney 2006}

\author[0000-0002-6510-0681]{Motohide Tamura}
\affiliation{Department of Astronomy, The University of Tokyo, 7-3-1, Hongo, Bunkyo-ku, Tokyo 113-0033, Japan}
\affiliation{Astrobiology Center of NINS, 2-21-1 Osawa, Mitaka, Tokyo 181-8588, Japan}
\affiliation{National Astronomical Observatory of Japan, 2-21-1 Osawa, Mitaka, Tokyo 181-8588, Japan}
    
\author[0000-0003-4018-2569]{Sebastien Vievard}
\affiliation{Subaru Telescope, National Astronomical Observatory of Japan, National Institutes of Natural Sciences, 650 North A`oh$\bar{o}$k$\bar{u}$ Place, Hilo, HI 96720, USA}



\begin{abstract}

Recent high-sensitivity observations reveal that accreting giant planets embedded in their parental circumstellar disks can emit H$\alpha$ at their final formation stages.
While the origin of such emission is not determined yet, magnetospheric accretion is currently a most plausible hypothesis.
In order to test this hypothesis further, we develop a simplified, but physical-based model and apply it to our observations taken toward HD 163296 with Subaru/SCExAO+VAMPIRES.
We specify under what conditions, embedded giant planets can undergo magnetospheric accretion and emit hydrogen lines.
We find that when stellar accretion rates are high, magnetospheric accretion becomes energetic enough to self-regulate the resulting emission.
On the other hand, if massive planets are embedded in disks with low accretion rates, earlier formation histories determine whether magnetospheric accretion occurs.
We explore two different origins of hydrogen emission lines (magnetospheric accretion flow heated by accretion-related processes vs planetary surfaces via accretion shock).
The corresponding relationships between the accretion and line luminosities dictate that emission from accretion flow achieves higher line flux than that from accretion shock 
and the flux decreases with increasing wavelengths (i.e., from H$\alpha$ to Pa$\beta$ and up to Br$\gamma$).
Our observations do not detect any point-like source emitting H$\alpha$ and are used to derive the 5$\sigma$ detection limit.
The observations are therefore not sensitive enough, and
reliable examination of our model becomes possible if observational sensitivity will be improved by a factor of ten or more.
Multi-band observations increase the possibility of efficiently detecting embedded giant planets and carefully determining the origin of hydrogen emission lines.

\end{abstract}



\section{Introduction} \label{sec:intro}

Understanding giant planet formation is fundamental in astrophysics and planetary science today.
NASA's {\it Kepler} mission and other astronomical observations reveal that giant planets orbit around their host star
with a wide range of orbital periods \citep[$\sim 0.02 - 7 \times 10^5$ days, e.g.,][]{2015ARA&A..53..409W}.
NASA's Juno mission is currently unveiling the origin and interior structure of Jupiter \citep[e.g.,][]{2017GeoRL..44.4649W}.
More recently, Europa, one of Jupiter's moons, has been selected as a target for exploring the potential for life on other worlds.
It is thus vital to understand how giant planets form out of circumstellar disks.

It has widely been accepted that planet-forming environments are dense and cold \citep[e.g.,][]{2011ARA&A..49...67W}.
Hence, observationally exploring growing (proto)planets that are deeply embedded in such environments are hard.
While this view still holds for most stages (e.g., core formation and initial gas accretion) of planet formation, 
recent high-spatial resolution and high-sensitivity observations have demonstrated that later (or final) stages of giant planet formation can be studied observationally 
\citep[e.g.,][]{2015ApJ...808L...3A,2018A&A...617A..44K,2018ApJ...863L...8W}.
This becomes possible because ongoing giant planet formation exhibits potentially detectable signatures 
\citep[e.g.,][]{2002ApJ...566L..97W,2015ApJ...799...16Z,2018ApJ...866...84A,2022A&A...657A..38M}. 
One famous example are nearly concentric gaps in both the gas and dust distributions of disks \citep[e.g.,][]{2015ApJ...808L...3A,2020ARA&A..58..483A}.
Discoveries of such gaps are a triumph for theory of planet formation as many theoretical studies predicted their presence due to disk-planet interaction
\citep[e.g.,][]{2002ApJ...566L..97W,2012ARA&A..50..211K}.

Another breakthrough achieved by recent observations, which is the topic of this work, 
are detections of H$\alpha$ emission coming from young giant planets orbiting around PDS 70 
\citep{2018A&A...617A..44K,2018A&A...617L...2M,2018ApJ...863L...8W,2019NatAs...3..749H}.
Similar detections have been claimed for other disks \citep[e.g., LkCa 15,][]{2015Natur.527..342S}.
However, robust confirmation of point-like sources as accreting giants is challenging 
because H$\alpha$ emission can also be caused by stellar light that is scattered by disks' inner edge.
In fact, both cases (emission from accreting planets and scattered stellar light from the inner edge) are possible for some targets \citep[e.g., ][]{2019ApJ...877L...3C,2022NatAs...6..751C}.
Therefore, careful vetting of such detections is necessary.
PDS 70 b and c survive such vetting and are recognized as bona fide accreting giant planets in the community today 
\citep[][]{2018A&A...617A..44K,2018A&A...617L...2M,2018ApJ...863L...8W,2019NatAs...3..749H,2019ApJ...877L..33C,2020AJ....159..222H,2020AJ....159..263W,2021AJ....161..244Z}.

The origin of H$\alpha$ emission from young giant planets is currently under active investigation \citep[e.g.,][]{2020arXiv201106608A,2020ApJ...902..126S,2022A&A...657A..38M}.
One leading hypothesis is that these planets undergo magnetospheric accretion \citep[e.g.,][]{2019ApJ...885L..29A,2019ApJ...885...94T,2021ApJ...923...27H},
as with the case for classical T Tauri stars \citep[CTTS, e.g.,][]{1991ApJ...370L..39K,2016ARA&A..54..135H}.
In this picture, accreting giant planets have sufficiently strong magnetic fields, so that the surrounding circumplanetary disks are truncated.
Gas accretion from the disks onto planets proceeds through planetary magnetospheres.
The emitting location of H$\alpha$ and hence its origin are still unclear; 
it may come from either accretion flow like CTTSs \citep{2019ApJ...885...94T} or accretion shock at planetary surfaces \citep{2019ApJ...885L..29A}.
Next steps are therefore to identify where (and how) the observed H$\alpha$ emission is produced by accreting giant planets 
and to quantify how common such emission is at the later stages of giant planet formation.

To this goal, we develop a simplified, but physical-based model 
to theoretically predict under what conditions, accreting giant planets emit (observable) hydrogen lines due to magnetospheric accretion.
In order to increase the sample size and to quantify the ubiquity of H$\alpha$ emission during giant planet formation,
we also conduct new observations targeting HD 163296 with Subaru/SCExAO+VAMPIRES.
We will show below that our theoretical calculations provide predictions for hydrogen emission lines,
while our observations are not sensitive enough;
reliable determination of the emission mechanism/location of H$\alpha$ from giant planets deeply embedded in their parental circumstellar disks
requires that observational sensitivity needs to be improved by at least a factor of ten.
The exact degree of improvement depends heavily on extinction of planet-forming regions, which is poorly constrained currently.
Feasibility of observational tests increases at longer wavelengths because the effect of extinction becomes less severe;
Pa$\beta$ and/or Br$\gamma$ lines would be better tracers of accretion processes for deeply embedded planets.
Ongoing and planned JWST observations will detect such lines \citep[e.g.,][]{2023ApJ...949L..36L}.

Our target, HD 163296, is a Herbig Ae/Be star surrounded by the gapped, circumstellar disk \citep[e.g.,][]{2007A&A...469..213I,2013A&A...557A.133D,2016PhRvL.117y1101I}.
It is located at $\sim 100$ pc away from Earth \citep{2016A&A...595A...1G,2023A&A...674A...1G},
and its mass and age are $\sim 2.3 ~ M_{\odot}$ and $\sim 5$ Myr, respectively \citep[e.g.,][]{1997A&A...324L..33V}.
This young stellar object is an ideal testbed due to the following four reasons:
1) meridional flows are detected by the $^{12}$CO $j=2-1$ emission \citep{2019Natur.574..378T},
which may be produced by the presence of giant planets;
2) the gas velocity kink is discovered by the $^{12}$CO $j=2-1$ emission \citep{2018ApJ...860L..13P}, 
which is now accepted as a reliable exoplanet detection method \citep{2019NatAs...3.1109P};
3) both gas and dust multiple gaps are observed in the disk, highly suggesting the presence of accreting, not-yet-directly-observed giant planets \citep{2016PhRvL.117y1101I}; and
4) a detection of a point source is reported via direct imaging \citep{2018MNRAS.479.1505G} 
while the follow-up observations do not verify its presence yet \citep{2019ApJ...875...38R}.
More recently, a localized kinematic structure has been reported in atomic carbon emission,
which spatially coincides with the innermost planet candidate \citep{2022ApJ...941L..24A}.
Table \ref{table1} summarizes the properties of giant planet candidates inferred from various observational signatures
with the fiducial values used in this work.

\begin{table*}
\begin{minipage}{17cm}
\centering
\caption{The properties of giant planet candidates embedded in the disk around HD 163296}
\label{table1}
{
\begin{tabular}{c|c|c|c|c|c}
\hline
Name           & Inferred method   & Planet position $r_{\rm p}$ (au)$^a$ & Planet mass $M_{\rm p}$ ($M_{\rm J}$)$^a$  & Planet radius $R_{\rm p}$ ($R_{\rm J}$)$^b$ & Reference \\ \hline \hline
Candidate 1 & Dust gap              & $\sim 60$                 &  $\sim 0.5-2$                        &                                                     & \citet{2016PhRvL.117y1101I}      \\
                    & Direct imaging      & $\sim 50$                 &  $\sim 6-7$                           &                                                     & \citet{2018MNRAS.479.1505G}   \\  \hline
Fiducial       &                               & $ 55$                       &  3.8                                      & 1.5                                               &                                                      \\  \hline \hline
Candidate 2 & Dust gap              & $\sim 100$               &  $\sim 0.05-0.3$                   &                                                     & \citet{2016PhRvL.117y1101I}      \\  
                    & Meridional flow     & $\sim 87$                  &  $\sim 0.5$                           &                                                     & \citet{2019Natur.574..378T}        \\  \hline
Fiducial       &                             & $ 94$                        &  0.28                                      & 1.5                                                &                                                     \\  \hline  \hline
Candidate 3 & Dust gap              & $\sim 160$               &  $\sim 0.15-0.5$                   &                                                      &  \citet{2016PhRvL.117y1101I} \\  
                    & Meridional flow     & $\sim 140$               &  $\sim 1$                              &                                                      & \citet{2019Natur.574..378T}   \\  \hline
Fiducial        &                               & $150$                       &  0.33                                         & 1.5                                                 &                                                     \\  \hline  \hline
Candidate 4 & Gas velocity kink  & $\sim 260$               &  $\sim 2$                              &                                                        & \citet{2018ApJ...860L..13P}  \\ 
                     & Meridional flow     & $\sim 237$              &  $\sim 2$                               &                                                     & \citet{2019Natur.574..378T}   \\  \hline
Fiducial        &                              & $ 249$                    &  2                                          & 1.5                                                 &                                                     \\  \hline  \hline
\end{tabular}

$^a$The fiducial values of $r_{\rm p}$ and $M_{\rm p}$ are obtained, by computing intermediate values of the given ranges.

$^b$ Both theoretical and observational studies suggest that the mass-radius relation becomes fairly flat for Jovian planets 
\citep[e.g.,][]{2012A&A...547A.112M,2017ApJ...834...17C}.
Hence, we adopt the constant value in this work.
}
\end{minipage}
\end{table*}

The plan of this paper is as follows.
In Section \ref{sec:mod}, we develop a simplified, but physical-based model to provide theoretical predictions of when and how accreting giant planets emit hydrogen lines.
In Section \ref{sec:data}, we summarize our observations and compare theoretical predictions with observational results.
In Section \ref{sec:disc}, we discuss assumptions adopted in our model and the limitation of our model.
Section \ref{sec:sum} is devoted to the summary of this work.

\section{Theoretical prediction} \label{sec:mod}

We provide theoretical predictions of when hydrogen lines can be emitted from young giant planets undergoing magnetospheric accretion.
The fundamental assumption of this work is that planetary magnetic fields are powered by accretion onto these planets.
We will show below that this assumption is reasonable for certain masses of planets and enables self-consistent calculations.

\subsection{Energy budget} \label{sec:mod_en}

When planets accrete the surrounding gas  with an accretion rate of $\dot{M}_{\rm p}$,
the total accretion luminosity ($L_{\rm acc}$) is given as
\begin{eqnarray}
\label{eq:L_acc}
L_{\rm acc} & =          &\frac{G M_{\rm p} \dot{M}_{\rm p}}{R_{\rm p}}  \\ \nonumber
                   & \simeq & 1.2 \times 10^{-4} L_{\odot} \left( \frac{M_{\rm p}}{10M_{\rm J}} \right) \left( \frac{\dot{M}_{\rm p}}{ 10^{-7} M_{\rm J} \mbox{ yr}^{-1} } \right) \left( \frac{R_{\rm p}}{1.5R_{\rm J} } \right)^{-1},
\end{eqnarray}
where $L_{\odot}$ is the solar luminosity, $M_{\rm p}$ and $R_{\rm p}$ are the planet mass and radius, and $M_{\rm J}$ and $R_{\rm J}$ are Jupiter's mass and radius, respectively.
Hereafter, we adopt $R_{\rm p} = 1.5 R_{\rm J}$ (Table \ref{table1});
the radius evolution of planets becomes minimal after the initial contraction of planetary envelopes with the Kelvin-Helmholtz timescale ends 
\citep[e.g.,][]{2000Icar..143....2B,2012A&A...547A.111M}.
In the subsequent stage, planets undergo the so-called disk-limited gas accretion \citep[e.g.,][]{2019ApJ...876L..32H},
and rotationally supported, circumplanetary disks should emerge due to the conservation of angular momentum.
Magnetospheric accretion may come into play at the disk-limited gas accretion stage.
In equation (\ref{eq:L_acc}), we adopt characteristic values for $M_{\rm p}$ and $\dot{M}_{\rm p}$ 
suggested for PDS 70 b/c as an example \citep[][references herein]{2021ApJ...923...27H};
these values vary in the following sections.

The presence of circumplanetary disks divides the accretion luminosity into two components: 
The luminosity coming from disks ($L_{\rm disk}= f_{\rm in} L_{\rm acc}$), 
and the one originating from energy liberated as disks' gas falls onto planets ($L_{\rm gas} = (1-f_{\rm in} )L_{\rm acc}$),
where $f_{\rm in}$ is the partition coefficient of accretion energy that is controlled by the location of disks' inner edge ($R_{\rm in}$).
When disks are heated predominantly by viscosity, $f_{\rm in}$ is written as \citep[e.g.,][]{1981ARA&A..19..137P}
\begin{equation}
\label{eq:f_in}
f_{\rm in} = \frac{3}{2} \left( \frac{R_{\rm p}}{R_{\rm in}} \right) \left[1 - \frac{2}{3} \left( \frac{R_{\rm p}}{R_{\rm in}} \right)^{1/2} \right].
\end{equation}
For simplicity, we adopt the above expression for $f_{\rm in}$;
if $R_{\rm in} = R_{\rm p}$, then $f_{\rm in}=1/2$.
In this paper, $L_{\rm disk}$ and $L_{\rm gas}$ are referred to as the disk and infall gas luminosities, respectively.

The infall gas luminosity can be further decomposed into two components when planets undergo magnetospheric accretion, that is, $R_{\rm in} >R_{\rm p}$;
for this case, infall gas is channeled by magnetic field lines of accreting planets, and it reaches planetary surfaces at nearly free-fall velocity (see Section \ref{eq:mag_flow}).
The gas radiates some energy both when it is in the magnetospheric flow and when it produces shock at planetary surfaces.
The corresponding luminosity can be written as $L_{\rm rad} = (1-f_{\rm L})L_{\rm gas}$,
where $f_{\rm L}$ is the partition coefficient of $L_{\rm gas}$.
It is important that all of the energy of the infall gas cannot radiate away, 
and hence some of the energy should be thermalized with the atmospheric gas of planets.
Accordingly, accreting giant planets are heated up by the gas coming from circumplanetary disks to some extent.
We label such luminosity as $ L_{\rm therm} = f_{\rm L} L_{\rm gas}$.
\citet{2020arXiv201106608A} estimate the value of $f_{\rm L}$ due to accretion shock at planetary surfaces and find that $f_{\rm L} \simeq 0.3-0.8$.

\begin{figure}
\begin{center}
\includegraphics[width=8.3cm]{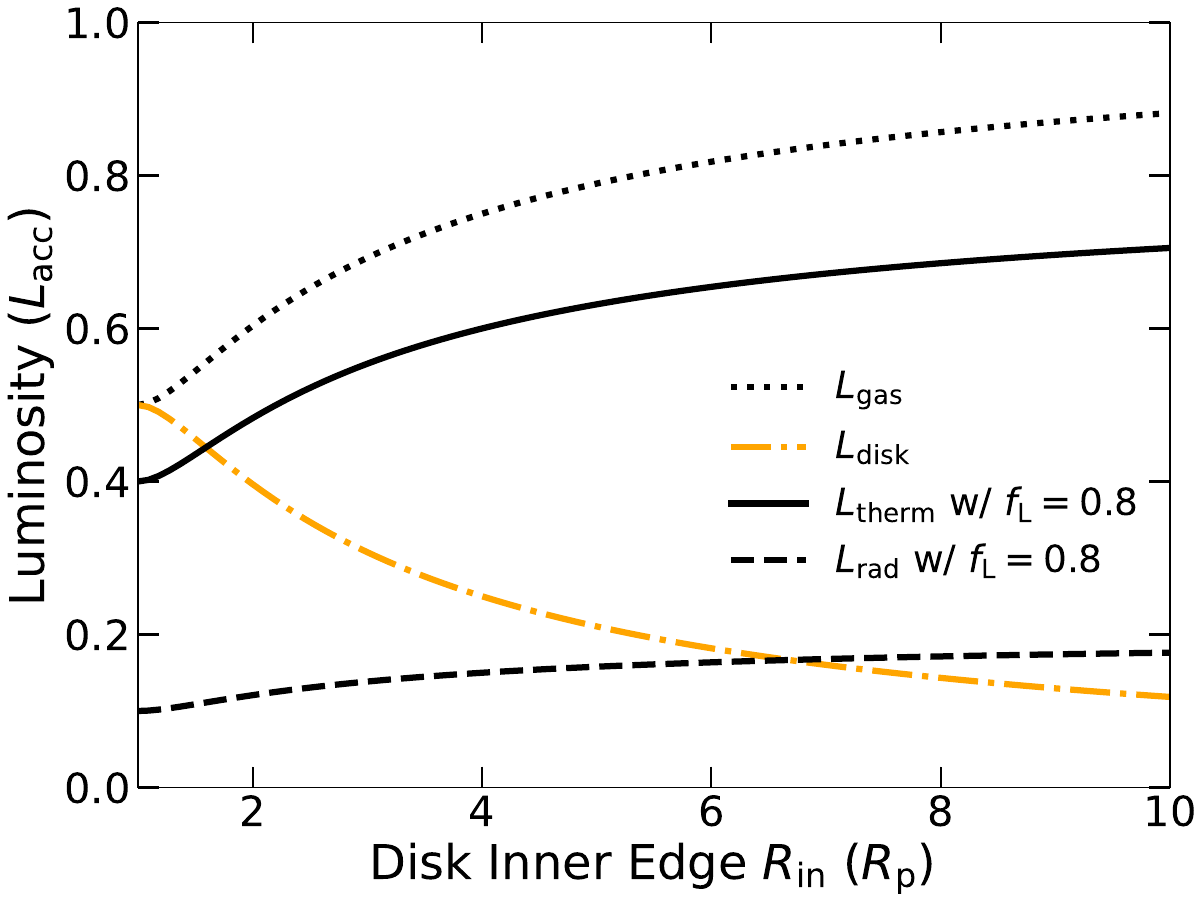}
\caption{The luminosities associated with accreting planets as a function of the disk inner edge.
When $R_{\rm in} > R_{\rm p}$, most of the accretion energy is carried by the infall gas (i.e., $L_{\rm gas}$).
Such gas can heat up the host, accreting planets via $L_{\rm therm}$.
As an example, $f_{\rm L}=0.8$ is adopted in this plot.
}
\label{fig1}
\end{center}
\end{figure}

In summary, the accretion luminosity can be written as
\begin{eqnarray}
L_{\rm acc} & = & L_{\rm disk} + L_{\rm gas}  \\ \nonumber
                   & = & L_{\rm disk} + L_{\rm rad} + L_{\rm therm}, 
\end{eqnarray}
where $L_{\rm disk}= f_{\rm in} L_{\rm acc}$, $L_{\rm rad} = (1-f_{\rm L})(1-f_{\rm in} )L_{\rm acc}$, and  $L_{\rm  therm} = f_{\rm L}(1-f_{\rm in} )L_{\rm acc}$.
Figure \ref{fig1} shows how each component of luminosities change as a function of $f_{\rm in}$ (or $R_{\rm in}$) and $f_{\rm L}$.
As expected, $L_{\rm gas}$ exceeds $L_{\rm disk}$ when $R_{\rm in} > R_{\rm p}$.
Also, $L_{\rm therm}$ can contribute to about up to 70 \% of $L_{\rm acc}$ when $f_{\rm L}=0.8$.
This indicates that the effective temperature of accreting planets is affected by accretion considerably.

In the following calculations, we adopt that $f_{\rm L}=0.5$ as it is an intermediate value.

\subsection{Effective temperature}

The effective temperature is one important quantity to characterize the properties of accreting planets.
In this work, it becomes the key parameter to estimate planetary magnetic fields.
We here compute the effective temperature of accreting giants, using $L_{\rm therm}$ discussed above.

Quantifying the effective temperature of young giant planets receives considerable attention in the literature.
This is because it may be used as a diagnostics to differentiate their formation mechanisms \citep[e.g.,][]{2007ApJ...655..541M,2012ApJ...745..174S};
the so-called hot- and cold-starts (i.e., large-sized planets with high temperatures and small-sized planets with low temperatures)
may be realized as the results of two completing planet formation scenarios: gravitational instability and core accretion, respectively.
The advent of direct imaging technique to search for young giant planets enables measurements of their luminosity,
and hence their size and temperature can be estimated \citep[e.g.,][]{2008Sci...322.1348M}.
This capability thus makes it possible to specify formation mechanisms of these planets.
However, no conclusive remark is made in the literature yet.

In this work, the effective temperature of accreting giants is computed as 
\begin{equation}
\label{eq:T_pe}
T_{\rm p,e}^4 = T_{\rm int}^4 + \frac{ L_{\rm therm} }{4 \pi \sigma_{\rm SB} R_{\rm p}^2},
\end{equation}
where $T_{\rm int}$ is the intrinsic temperature of planets, and $\sigma_{\rm SB}$ is  the Stefan-Boltzmann constant.
The above equation assumes that accreted gas with luminosity of $L_{\rm therm}$ is thermalized over the entire surface of planets.
This is most conservative because $T_{\rm p,e}$ takes the lowest value.
Inclusion of $L_{\rm therm}$ in equation (\ref{eq:T_pe}) corresponds to the so-called warm start as some of accreted gas heats up planets.
Reliable calculations of $T_{\rm int}$ require tracking of planet formation histories from the beginning as done by \citet[e.g.,][]{2012A&A...547A.112M},
which is beyond the scope of this work.
We therefore assume that $T_{\rm int}=700$ K, following \citet{2012ApJ...745..174S}.

Heating by $L_{\rm therm}$ leads to the following, planet surface temperature:
\begin{eqnarray}
\label{eq:T_therm}
T_{\rm therm} & \equiv & \left(  \frac{ L_{\rm therm} }{4 \pi \sigma_{\rm SB} R_{\rm p}^2} \right)^{1/4} \\ \nonumber
                       & \simeq & 1.5 \times 10^{3} \mbox{K}  \left( \frac{f_{\rm L}}{1/2} \right)^{1/4} \left( \frac{1-f_{\rm in}}{1/2} \right)^{1/4} \\ \nonumber
                       & \times & \left( \frac{M_{\rm p}}{10M_{\rm J}} \right)^{1/4} \left( \frac{\dot{M}_{\rm p}}{ 10^{-7} M_{\rm J} \mbox{ yr}^{-1} } \right)^{1/4} \left( \frac{R_{\rm p}}{1.5R_{\rm J} } \right)^{-3/4}.
\end{eqnarray}

We will use below both $T_{\rm int}$ and $T_{\rm therm}$ to explore what strength of planetary magnetic fields is produced by these temperatures.

\subsection{Planetary magnetic fields} \label{sec:pla_mag}

Planetary magnetic fields are generated by dynamo activities operating in electrically conducting interiors,
where convective motion occurs.
Precise determination of the field strength is hard.
To make the problem tractable,
we here use a scaling law available in the literature and estimate the strength of planetary magnetic fields.

In principle, the ultimate source of energy to invoke dynamo activities is the thermodynamic energy available in the interior of planets;
the energy is converted to magnetic energy, and thermal flux is maintained against ohmic dissipation.
\citet{2009Natur.457..167C} adopt this principle and derive a scaling law, which is written as
\begin{equation}
\label{eq:scale_law2}
\frac{\langle B \rangle^2}{8 \pi} = c f_{\rm ohm} \langle \rho \rangle^{1/3} (F q)^{2/3},
\end{equation}
where $\langle B \rangle$ is the mean magnetic field on the dynamo surface, $c$ is a constant of proportionality, 
$f_{\rm ohm} \simeq 1 $ is the ratio of ohmic dissipation to total dissipation, 
$\langle \rho \rangle$ is the mean bulk density of planets where the field is generated, 
$F=0.35$ is the efficiency factor of converting thermal energy to magnetic energy,
$q= \sigma_{\rm SB}T^4_{\rm p,e}$.
A value  of $c \simeq1.1$ is obtained by adopting the typical values of Jupiter ($B_{\rm p, s}=10$ G and $q=5.4 \times 10^3$ erg s$^{-1}$ cm$^{-2}$)
and assuming that $\langle B \rangle/B_{\rm p, s} \simeq 7$,
where $B_{\rm p, s}$ is the magnetic field strength at planetary surfaces.
Remarkably, \citet{2009Natur.457..167C} show that the law successfully reproduces magnetic fields of objects reasonably well 
from solar system planets (e.g., Earth and Jupiter) up to rapidly rotating stars such as CTTSs.

We use the above scaling law to compute the magnetic field strength of accreting giants.
Combing equations (\ref{eq:T_pe}) and (\ref{eq:scale_law2}), 
magnetic fields of accreting giant planets can be given as
\begin{eqnarray}
\label{eq:B_ps}
B_{\rm p, s}  & \simeq &  \frac{ \langle B \rangle }{7}  \\ \nonumber
                    & =          &  \frac{6^{1/6}}{7} (8 \pi)^{1/3} (c f_{\rm ohm}  )^{1/2}  (F \sigma_{\rm SB})^{1/3}M_{\rm p}^{1/6} R_{\rm p}^{-1/2} T_{\rm p,e}^{4/3}.
\end{eqnarray}
When two limits are considered for $T_{\rm p,e}$, $B_{\rm p, s}$ is rewritten as 
\begin{eqnarray}
\label{eq:B_ps1}
B_{\rm p, s}^{\rm int} 
                        &  \simeq & 1.6 \times 10^2  \mbox{G}   \\ \nonumber
                        & \times &    \left( \frac{M_{\rm p}}{10M_{\rm J}} \right)^{1/6}  \left( \frac{R_{\rm p}}{1.5R_{\rm J} } \right)^{-1/2} \left( \frac{T_{\rm int}}{700 {\rm K}} \right)^{4/3}.
\end{eqnarray}
for the case that $T_{\rm p,e} \simeq T_{\rm int}$, and
\begin{eqnarray}
\label{eq:B_ps2}
B_{\rm p, s}^{\rm therm} 
                         &  \simeq & 3.3 \times 10^2  \mbox{G}   \left( \frac{f_{\rm L}}{1/2} \right)^{1/3} \left( \frac{1 - f_{\rm in}}{1/2} \right)^{1/3}  \\ \nonumber
                        & \times &    \left( \frac{M_{\rm p}}{10M_{\rm J}} \right)^{1/2} \left( \frac{\dot{M}_{\rm p}}{ 10^{-7} M_{\rm J} \mbox{ yr}^{-1} } \right)^{1/3}
                         \left( \frac{R_{\rm p}}{1.5R_{\rm J} } \right)^{-3/2}.
\end{eqnarray}
for the case that $T_{\rm p,e} \simeq T_{\rm therm}$.

These calculations indicate that when $T_{\rm p,e} \simeq T_{\rm int}$,
$T_{\rm int}$ becomes the fundamental parameter to determine $B_{\rm p,s}$ (equation (\ref{eq:B_ps1}));
equivalently, earlier formation histories dictate whether or not magnetospheric accretion occurs.
On the other hand, when $T_{\rm p,e} \simeq T_{\rm therm}$,
all the key quantities (e.g., $T_{\rm p,e}$, $B_{\rm p,s}$, and $\dot{M}_{\rm p}$) can be computed self-consistently 
(see equations (\ref{eq:T_therm}) and (\ref{eq:B_ps2}), and also see equation (\ref{eq:mdot_p_therm}) as discussed below).
This essentially infers that disk-limited gas accretion can become energetic enough 
that physical parameters are self-regulated by the corresponding heating;
if magnetospheric accretion operates in such a gas accretion stage,
the resulting observables (e.g., hydrogen emission lines) serve as a direct probe of the stage.

In the following sections, we consider two limiting cases: $T_{\rm p,e} \simeq T_{\rm int}$ and $T_{\rm p,e} \simeq T_{\rm therm}$,
and explore under what conditions, giant planets undergo magnetospheric accretion and when (observable) hydrogen lines can be emitted.

\subsection{Magnetospheric accretion} \label{sec:mag_sph}

Magnetospheric accretion is currently a leading hypothesis to explain the observed H$\alpha$ emission from PDS 70 b/c 
\citep[e.g.,][]{2019ApJ...885L..29A,2019ApJ...885...94T,2021ApJ...923...27H}.
This accretion mode takes action when magnetic fields of planets are strong enough that 
accreting circumplanetary disks are truncated \citep[e.g.,][]{2015ApJ...799...16Z,2018AJ....155..178B,2021ApJ...923...27H}.
In this section, we determine when such a condition is met.

Circumplanetary disks are truncated
when the magnetic pressure ($B_{\rm p}^2/8\pi$) of host planets exceeds the ram pressure of accreting disks \citep{1979ApJ...232..259G}.
Mathematically, it is written as
\begin{equation}
\label{eq:R_T1}
\frac{B_{\rm p}^2}{8 \pi}  = f_{\rm ram}  \rho_{\rm ram} v_{\rm Kep}^2,
\end{equation}
where $v_{\rm Kep} = \sqrt{GM_{\rm p}/R}$ is the Keplerian velocity around planets, 
$R$ is the distance at the disk midplane measured from the planet center,
and $\rho_{\rm ram} \sim \dot{M}_{\rm p} / (4 \pi R^2 v_{\rm Kep})$ is the ram pressure of disks. 
A value of $f_{\rm ram} = 1/\sqrt{2}$ is adopted, following \citet{1979ApJ...232..259G}.
We also assume that the magnetic field ($B_{\rm p}$) of planets may be represented well as dipole, that is,
\begin{equation}
\label{eq:B_p}
B_{\rm p} (r) = B_{\rm p,s}(R/R_{\rm p})^{-3}.
\end{equation}

From equations (\ref{eq:B_ps}), (\ref{eq:R_T1}), and (\ref{eq:B_p}), 
one can derive a relationship between $M_{\rm p}$ and $\dot{M}_{\rm p}$ for a given value of $R$.
Considering the two limits for $T_{\rm p,e}$, 
$\dot{M}_{\rm p}$ is given as
\begin{eqnarray}
\label{eq:mdot_p_int}
\dot{M}_{\rm p}^{\rm int} & \simeq & 2.4 \times 10^{-8} M_{\rm J} \mbox{ yr}^{-1} 
                                                                           \left( \frac{f_{\rm ram}}{1/ \sqrt{2}} \right)^{-1}  \\ \nonumber
                                       & \times   & \left( \frac{M_{\rm p}}{10M_{\rm J}} \right)^{-1/6}  \left( \frac{R_{\rm p}}{1.5R_{\rm J} } \right)^{3/2} \left( \frac{T_{\rm int}}{700 {\rm K}} \right)^{8/3} 
                                                      \left( \frac{ R_{\rm in}/R_{\rm p} }{4} \right)^{-7/2} 
 \end{eqnarray}
for the case that $T_{\rm p,e} \simeq T_{\rm int}$, and
\begin{eqnarray}
\label{eq:mdot_p_therm}
\dot{M}_{\rm p}^{\rm therm} & \simeq & 2.5 \times 10^{-7} M_{\rm J} \mbox{ yr}^{-1} \left( \frac{f_{\rm ram}}{1/ \sqrt{2}} \right)^{-3} \left( \frac{f_{\rm L}}{1/2} \right)^{2} \\ \nonumber
                                            & \times &        \left( \frac{1 - f_{\rm in}}{3/4} \right)^{2}    \left( \frac{M_{\rm p}}{10M_{\rm J}} \right)^{3/2}  \left( \frac{R_{\rm p}}{1.5R_{\rm J} } \right)^{-3/2}  \\ \nonumber
                                            & \times &         \left( \frac{ R_{\rm in}/R_{\rm p} }{4} \right)^{-21/2} 
\end{eqnarray}
for the case that $T_{\rm p,e} \simeq T_{\rm therm}$.
Note that in the above calculations, we set that $R=R_{\rm in} = 4 R_{\rm p}$, that is, $f_{\rm in}=1/4$;
equivalently, disks' truncation radius due to planetary magnetospheres corresponds to their inner edge.
Also, $\dot{M}_{\rm p}^{\rm therm} $ is computed self-consistently, and hence it becomes a function of $M_{\rm p}$ and $R_{\rm p}$ (equation (\ref{eq:mdot_p_therm})).

We are now in a position to determine under what condition, planetary magnetic fields become strong enough to truncate circumplanetary disks.
To proceed, we examine all the quantities ($T_{\rm p,e}$, $B_{\rm p,s}$, $\dot{M}_{\rm p}$, and $L_{\rm acc}$) considered so far.
As discussed above, $T_{\rm int}$ is the fundamental parameter for the case that $T_{\rm p,e} \simeq T_{\rm int}$,
while these quantities are all computed self-consistently for the case that $T_{\rm p,e} \simeq T_{\rm therm}$.

We first summarize relevant equations.
For the case that $T_{\rm p,e} \simeq T_{\rm int}$,
$T_{\rm int} = 700$ K, $B_{\rm p,s}$ is described by equation (\ref{eq:B_ps1}), $\dot{M}_{\rm p}$ is given by equation (\ref{eq:mdot_p_int}),
and $L_{\rm acc}$ is written as
\begin{eqnarray}
L_{\rm acc}^{\rm int}  & \simeq & 4.4 \times 10^{-5} L_{\odot}    \left( \frac{f_{\rm ram}}{1/ \sqrt{2}} \right)^{-1}  \\ \nonumber
                                       & \times   & \left( \frac{M_{\rm p}}{10M_{\rm J}} \right)^{5/6}  \left( \frac{R_{\rm p}}{1.5R_{\rm J} } \right)^{1/2} \left( \frac{T_{\rm int}}{700 {\rm K}} \right)^{8/3} \\ \nonumber
                                        & \times   &     \left( \frac{ R_{\rm in}/R_{\rm p} }{4} \right)^{-7/2}. 
\end{eqnarray}
For the case that $T_{\rm p,e} \simeq T_{\rm therm}$,
\begin{eqnarray}
\label{eq:Tpe_therm}
T_{\rm therm} & \simeq & 1.7 \times 10^{3} \mbox{K}  \left( \frac{f_{\rm ram}}{1/ \sqrt{2}} \right)^{-3/4} \left( \frac{f_{\rm L}}{1/2} \right)^{3/4}    \\ \nonumber
                       & \times &  \left( \frac{1-f_{\rm in}}{3/4} \right)^{3/4} \left( \frac{M_{\rm p}}{10M_{\rm J}} \right)^{5/8} \left( \frac{R_{\rm p}}{1.5R_{\rm J} } \right)^{-9/8} \\ \nonumber
                      & \times &  \left( \frac{ R_{\rm in}/R_{\rm p} }{4} \right)^{-21/8}, 
\end{eqnarray}
where equations (\ref{eq:T_therm}) and (\ref{eq:mdot_p_therm}) are used,
\begin{eqnarray}
\label{eq:Bps_therm}
B_{\rm p, s}^{\rm therm}    &  \simeq & 5.1 \times 10^2  \mbox{G}  \left( \frac{f_{\rm ram}}{1/ \sqrt{2}} \right)^{-1}  \left( \frac{f_{\rm L}}{1/2} \right)  \\ \nonumber
                                          & \times &   \left( \frac{1 - f_{\rm in}}{3/4} \right)  \left( \frac{M_{\rm p}}{10M_{\rm J}} \right)   \left( \frac{R_{\rm p}}{1.5R_{\rm J} } \right)^{-2}  \\ \nonumber
                                         & \times &     \left( \frac{ R_{\rm in}/R_{\rm p} }{4} \right)^{-21/6},
\end{eqnarray}
where equations (\ref{eq:B_ps2}) and (\ref{eq:mdot_p_therm}) are used, 
$\dot{M}_{\rm p}^{\rm therm}$ is given by equation (\ref{eq:mdot_p_therm}), and
\begin{eqnarray}
L_{\rm acc}^{\rm therm}    &  \simeq & 4.7 \times 10^{-4}  L_{\odot}  \left( \frac{f_{\rm ram}}{1/ \sqrt{2}} \right)^{-3}  \left( \frac{f_{\rm L}}{1/2} \right)^2   \\ \nonumber
                                          & \times &  \left( \frac{1 - f_{\rm in}}{3/4} \right)^2 \left( \frac{M_{\rm p}}{10M_{\rm J}} \right)^{5/2}   \left( \frac{R_{\rm p}}{1.5R_{\rm J} } \right)^{-5/2} \\ \nonumber
                                          & \times & \left( \frac{ R_{\rm in}/R_{\rm p} }{4} \right)^{-21/2},
\end{eqnarray}
where equations (\ref{eq:L_acc}) and (\ref{eq:mdot_p_therm}) are used.

\begin{figure*}
\begin{minipage}{17cm}
\begin{center}
\includegraphics[width=8.3cm]{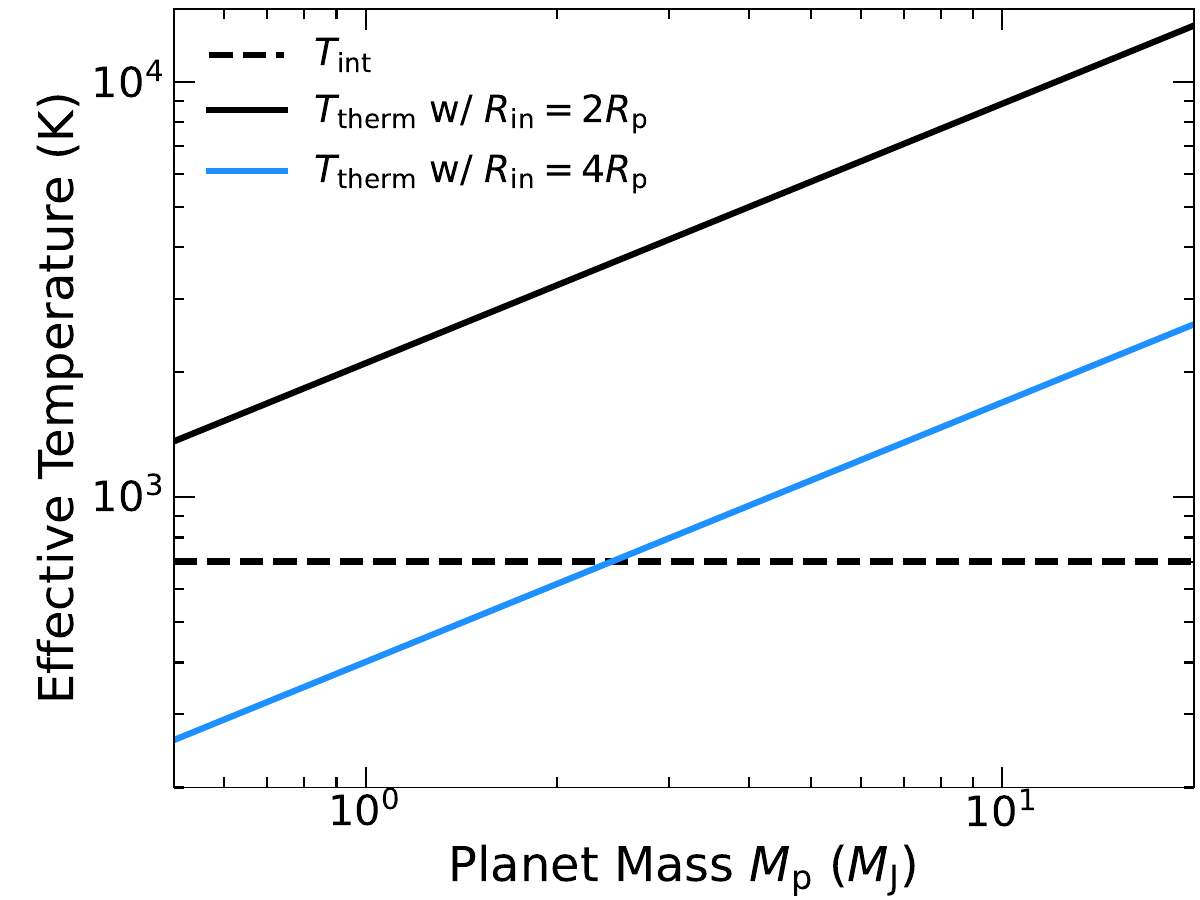}
\includegraphics[width=8.3cm]{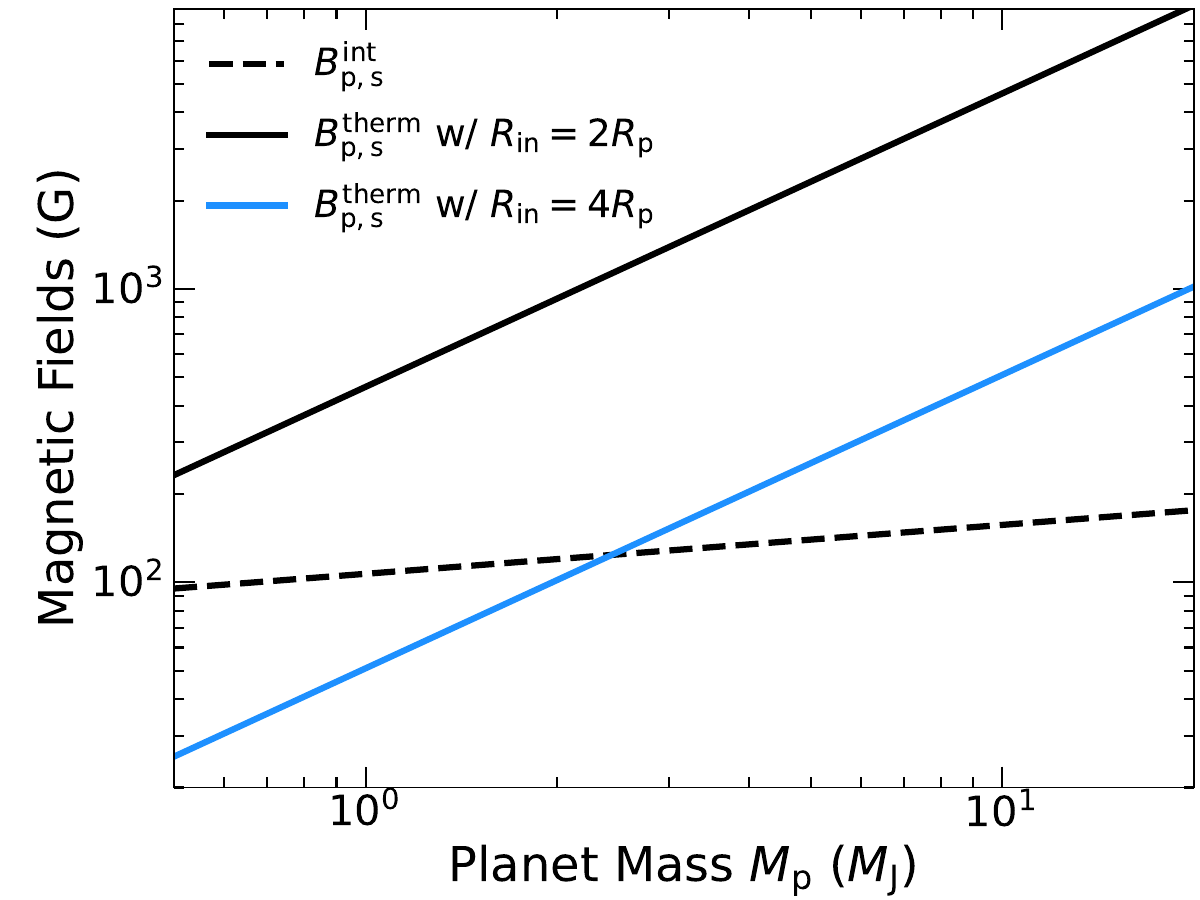}
\includegraphics[width=8.3cm]{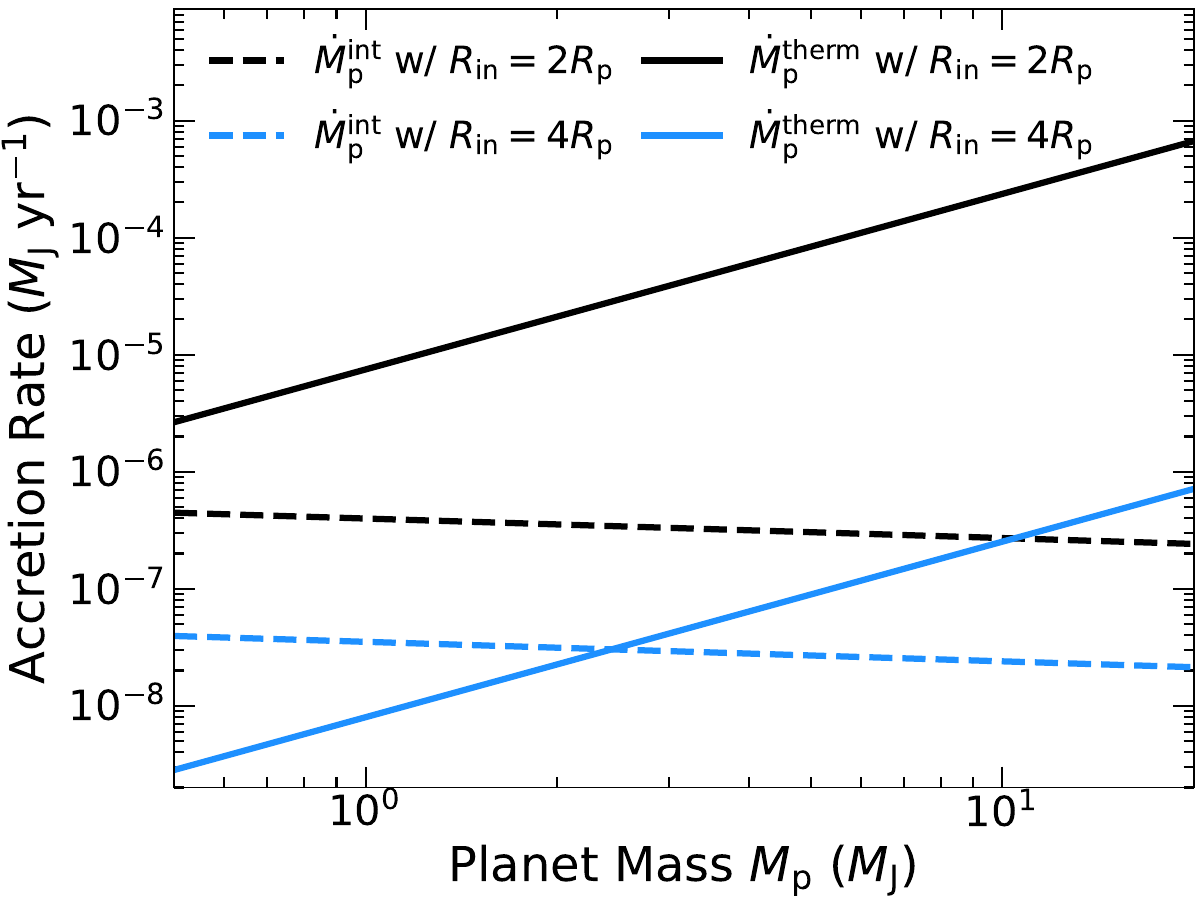}
\includegraphics[width=8.3cm]{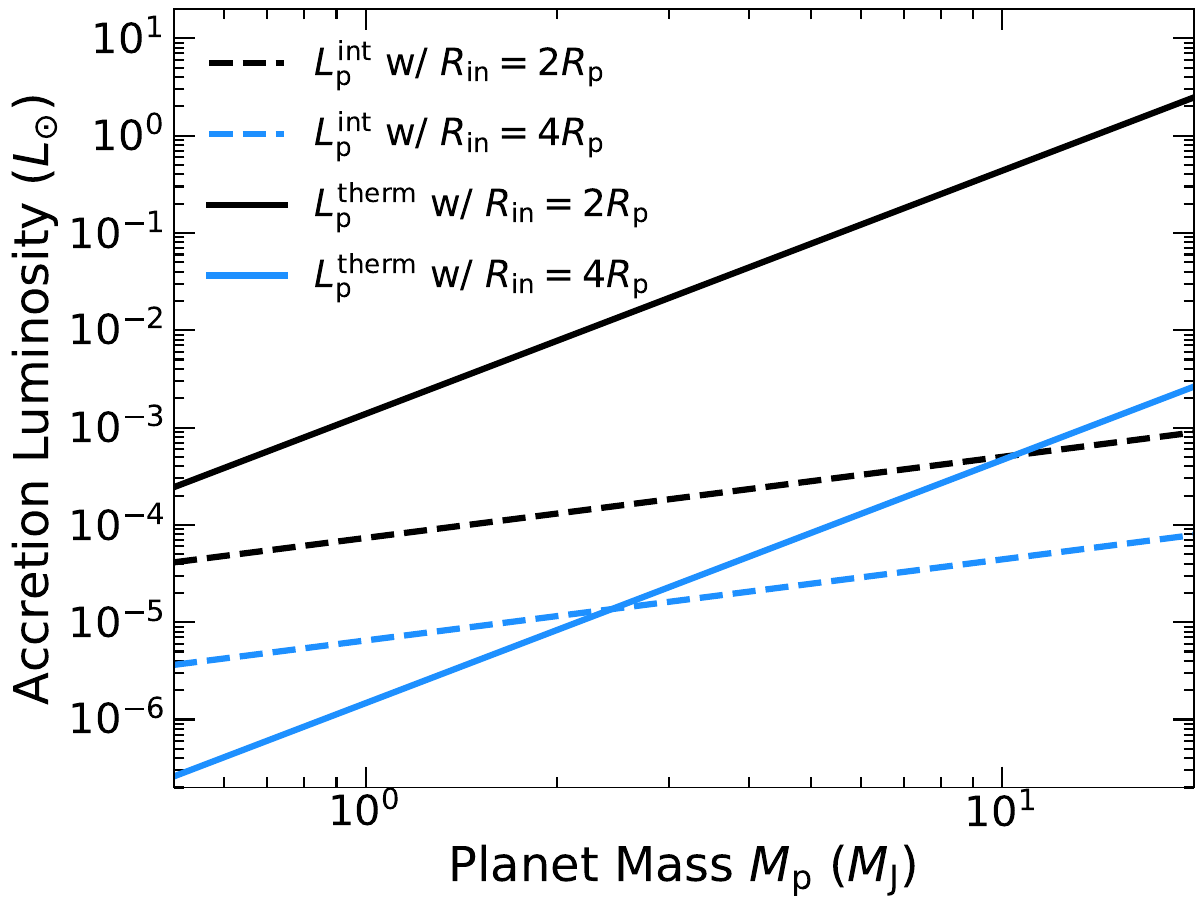}
\caption{The computed values of $T_{\rm p,e}$, $B_{\rm p,s}$, $\dot{M}_{\rm p}$, and $L_{\rm acc}$ as a function of $M_{\rm p}$ and $R_{\rm in}$ on
the top left, top right, bottom left and bottom right, respectively.
All the quantities are calculated at $R=R_{\rm in}$, 
and two values of $R_{\rm in}$ are adopted: $R_{\rm in}= 2R_{\rm p}(=3R_{\rm J})$, and $4R_{\rm p}(=6R_{\rm J})$.
In the case that $T_{\rm p,e} \simeq T_{\rm int}$, $T_{\rm int}$ is the fundamental parameter for $B_{\rm p,s}$,
and disk truncation by planetary magnetic fields becomes possible when the accretion rate and luminosity are lower than the dashed lines.
In the case that $T_{\rm p,e} \simeq T_{\rm therm}$, the solutions are obtained self-consistently,
and the solid lines represent the condition required for disk truncation.}
\label{fig2}
\end{center}
\end{minipage}
\end{figure*}

We then explore how these quantities behave as a function of $M_{\rm p}$ for given values of $R=R_{\rm in}$.
Figure \ref{fig2} shows the results.
In the plot, two values of $R_{\rm in}$ are considered: $R_{\rm in}= 2R_{\rm p}(=3R_{\rm J})$, and $4R_{\rm p}(=6R_{\rm J})$.
It is obvious that for the case that $T_{\rm p,e} \simeq T_{\rm int}$,
both $T_{\rm int}$ and $B_{\rm p,s}^{\rm int}$ are independent of $R_{\rm in}$ (see the dashed lines on the two top panels);
for $B_{\rm p,s}^{\rm int}$, it becomes a weak function of $M_{\rm p}$.
This is simply because planetary magnetic fields are regulated mainly by the effective temperature (see equation (\ref{eq:B_ps})).
For the case that $T_{\rm p,e} \simeq T_{\rm therm}$, 
both $T_{\rm therm}$ and $B_{\rm p,s}^{\rm therm}$ become a decreasing function of $R_{\rm in}$ and an increasing function of $M_{\rm p}$.
This is the direct outcome that these solutions are obtained self-consistently;
a small value of $R_{\rm in}$ means a small truncation radius of disks.
Such a situation tends to occur when the accretion rate is high.
To maintain disk truncation against the resulting high ram pressure, magnetic fields need to be strong, which in turn requires high effective temperatures.
For the $M_{\rm p}$ dependence, it can be understood as follows;
massive planets have deep gravitational potential, which leads to high ram pressure.
In order to prevent disks' inner edge from clashing onto planetary surfaces, high magnetic fields and hence high effective temperatures are required.

The behaviors of $\dot{M}_{\rm p}$ and $L_{\rm acc}$ are explained in a similar way (the two bottom panels).
Note that for the case that $T_{\rm p,e} \simeq T_{\rm int}$, the solutions are gained for a given value of $T_{\rm int}$.
Accordingly, these solutions should be viewed as an upper limit;
disk truncation can be achieved when the accretion rate is lower than the dashed lines (the bottom left panel of Figure \ref{fig2}).
Also, the negative and weak dependences on $M_{\rm p}$ arise due to a constant $T_{\rm int}$ for $\dot{M}_{\rm p}$ and $L_{\rm acc}$, respectively.
On the other hand, $\dot{M}_{\rm p}^{\rm therm}$ and $L_{\rm acc}^{\rm therm}$ are self-consistent solutions,
and hence the resulting values (denoted by the solid lines) are required to establish disk truncation due to magnetospheric accretion.

Thus, planetary magnetic fields become strong enough to truncate circumplanetary disks
when the disk accretion rate is lower than $\dot{M}_{\rm p}^{\rm int}$ for the case that $T_{\rm p,e} \simeq T_{\rm int}$
and when it becomes comparable to $\dot{M}_{\rm p}^{\rm therm}$ for the case that $T_{\rm p,e} \simeq T_{\rm therm}$.
It should be pointed out that these constraints are obtained as a function of $M_{\rm p}$ for given values of $R_{\rm in}$ in this section.
In the following sections, we use the solutions derived here to gain further constraints on $R_{\rm in}$

\subsection{Magnetospheric flow} \label{eq:mag_flow}

Gas in magnetospheric flow is known to be heated to $\sim 10^4$ K for CTTSs,
from which H$\alpha$ is emitted \citep[e.g.,][]{2001ApJ...550..944M}.
While the origin of the heat source is still unclear \citep[e.g.,][]{2016ARA&A..54..135H},
it should stem ultimately from magnetic fields of accreting objects and/or accretion energy.
For young giant planets, 
such heating should be attributed predominantly to accretion energy 
when disk-limited gas accretion results in high accretion rates;
as discussed in Section \ref{sec:pla_mag}, 
energetics of accretion processes can be self-regulated by the accompanying heating for the case that $T_{\rm p,e} \simeq T_{\rm therm}$.
We here consider such a case and derive a constraint on $R_{\rm in}$.

We first explore the properties of gas in magnetospheric flow.
Magnetospheric flow carries the following flux of energy when disks are truncated at $R=R_{\rm in}$ \citep[e.g.,][]{1998ApJ...509..802C}:
\begin{equation}
\label{eq:gas_kin}
\frac{1}{2} \dot{M}_{\rm p} v_{\rm sh}^2= \frac{G M_{\rm p} \dot{M}_{\rm p}}{R_{\rm p}} \left( 1 - \frac{R_{\rm p}}{R_{\rm in}} \right),
\end{equation}
where 
\begin{eqnarray}
v_{\rm sh}  & = & \sqrt{ \frac{2G M_{\rm p}}{R_{\rm p}} \left( 1 - \frac{R_{\rm p}}{R_{\rm in}} \right) } \\ \nonumber
                & \simeq & 1.3 \times 10^{2} \mbox{km s}^{-1} \left( \frac{ f_{\rm T} }{3/4} \right)^{1/2}
                    \left( \frac{M_{\rm p}}{10M_{\rm J}} \right)^{1/2}  \left( \frac{R_{\rm p}}{1.5R_{\rm J} } \right)^{-1/2}
\end{eqnarray}
is the velocity of accreted gas at planetary surfaces, and 
\begin{equation}
f_{\rm T} = 1 - \frac{R_{\rm p}}{R_{\rm in}}.
\end{equation}
In the above equation, we adopt that $R_{\rm p}/R_{\rm in}=1/4$, that is, $f_{\rm T} =3/4$.
The sound speed of gas at planetary surfaces with $T_{\rm p,e}$ of a few $10^3$ K (see Figure \ref{fig2}) becomes a few km s$^{-1}$
and is much smaller than $v_{\rm sh}$.
Therefore, shocks are produced at planetary surfaces when accreted gas arrives there.
Using the strong-shock approximation, the shock temperature ($T_{\rm sh}$) is given as 
\begin{eqnarray}
T_{\rm sh} & =          & \frac{3}{16} \frac{\mu m_{\rm H}}{k_{\rm B}} v_{\rm sh}^2 \\ \nonumber
                & \simeq &      4.0 \times 10^5 {\rm K}  \left( \frac{ \mu }{1} \right)  \left( \frac{ f_{\rm T} }{3/4} \right) \left( \frac{M_{\rm p}}{10M_{\rm J}} \right) 
                 \left( \frac{R_{\rm p}}{1.5R_{\rm J} } \right),
\end{eqnarray}
where $\mu$ is the mean molecular weight of accreted gas, $m_{\rm H}$ is the mass of hydrogen nucleons, and $k_{\rm B}$ is the Boltzmann constant.
The value of $\mu$ varies from $\sim 0.53$ to $\sim 1.28$ for ionized gas to neutral one at solar abundance.
Previous studies confirm that $T_{\rm sh}$ is high enough to both dissociate molecular hydrogen and ionize atomic hydrogen,
which leads to hydrogen line emission including H$\alpha$ \citep[e.g.,][]{2020arXiv201106608A}.

The generation of shocks at planetary surfaces is very likely and may be a most plausible explanation for PDS 70 b/c, as discussed above.
Nonetheless, it may be interesting to estimate the temperature of gas in magnetospheric flow 
and examine whether H$\alpha$ emission is possible from the flow, as with the case for CTTSs.
One conservative estimate of the flow temperature may be obtained,
assuming that the kinetic energy carried by magnetospheric flow completely dissipates before shocks occur
and that the emission would behave like blackbody at all the wavelengths.
The resulting flow temperature ($T_{\rm flow, BB}$) is computed as
\begin{equation}
\label{eq:T_flow}
4 \pi R_{\rm p}^2 f_{\rm fill} \sigma_{\rm SB} T_{\rm flow, BB}^4 = \frac{1}{2} \dot{M}_{\rm p} v_{\rm sh}^2,
\end{equation}
where $f_{\rm fill}$ is the so-called filling factor and represents a fraction of the planet surface area, 
at which magnetospheric flow arrives from the inner edge of circumplanetary disks.
Since equation (\ref{eq:T_flow}) is a function of $\dot{M}_{\rm p}$,
we consider both the cases that $T_{\rm p,e} \simeq T_{\rm int}$ and $T_{\rm p,e} \simeq T_{\rm therm}$ as done in Section \ref{sec:mag_sph}:
\begin{eqnarray}
T_{\rm flow, BB}^{\rm int} & \simeq  & 3.5 \times 10^{3} {\rm K} \left( \frac{f_{\rm fill}}{10^{-2}} \right)^{-1/4} \left( \frac{f_{\rm ram}}{1/ \sqrt{2}} \right)^{-1/4} \left( \frac{ f_{\rm T} }{3/4} \right)^{1/4} \\ \nonumber
                                       & \times   & \left( \frac{M_{\rm p}}{10M_{\rm J}} \right)^{5/24}  \left( \frac{R_{\rm p}}{1.5R_{\rm J} } \right)^{-3/8}  \left( \frac{T_{\rm int}}{700 {\rm K}} \right)^{2/3} \\ \nonumber
                                      & \times   &    \left( \frac{ R_{\rm in}/R_{\rm p} }{4} \right)^{-7/8} 
\end{eqnarray}
for the case that $T_{\rm p,e} \simeq T_{\rm int}$, 
where equation (\ref{eq:mdot_p_int}) is used, and 
\begin{eqnarray}
T_{\rm flow, BB}^{\rm therm} & \simeq  & 
                                  6.4 \times 10^{3} {\rm K} \left( \frac{f_{\rm fill}}{10^{-2}} \right)^{-1/4}  \left( \frac{f_{\rm ram}}{1/ \sqrt{2}} \right)^{-3/4} \left( \frac{f_{\rm L}}{1/2} \right)^{1/2} \\ \nonumber
                         & \times &    \left( \frac{1 - f_{\rm in}}{3/4} \right)^{1/2}  \left( \frac{ f_{\rm T} }{3/4} \right)^{1/4}   \left( \frac{M_{\rm p}}{10M_{\rm J}} \right)^{5/8}  \left( \frac{R_{\rm p}}{1.5R_{\rm J} } \right)^{-9/8} \\ \nonumber
                         & \times &  \left( \frac{ R_{\rm in}/R_{\rm p} }{4} \right)^{-21/8} \end{eqnarray}
for the case that $T_{\rm p,e} \simeq T_{\rm therm}$, 
where equation (\ref{eq:mdot_p_therm}) is used.

One recognizes that $T_{\rm flow, BB}$ becomes lower than $10^4$ K for both cases.
This implies that H$\alpha$ emission from magnetospheric flow would be unlikely for accreting giant planets if the emission behaves like blackbody.
The possibility of deviation from blackbody would be very likely, however;
the column density of accretion flow may not be high enough to achieve blackbody radiation at all the wavelengths.
Instead, the flow may be optically thick only for certain line emission such as H$\alpha$.
In fact, H$\alpha$ emission is known to be optically thick at the gas number density of  $n_{\rm H} > 10^{12}$ cm$^{-3}$ at a temperature of 8000 K for accretion flow onto CTTSs. 
\citep[e.g.,][]{1995MNRAS.272...41S,2015ApJ...799...16Z}. 
We find that a similar situation would be possible for accretion flow around accreting giants as
\begin{eqnarray}
n_{\rm H}^{\rm int} & = & \frac{\dot{M}_{\rm p}^{\rm int}}{ 4 \pi R_{\rm p}^2 f_{\rm fill} m_{\rm H} v_{\rm sh}} \\ \nonumber
                & \simeq & 4.5 \times 10^{12} {\rm cm}^{-3} \left( \frac{f_{\rm fill}}{10^{-2}} \right)^{-1}  \left( \frac{f_{\rm ram}}{1/ \sqrt{2}} \right)^{-1} \left( \frac{ f_{\rm T} }{3/4} \right)^{-1/2} \\ \nonumber
                                       & \times   & \left( \frac{M_{\rm p}}{10M_{\rm J}} \right)^{-2/3}   \left( \frac{T_{\rm int}}{700 {\rm K}} \right)^{8/3}   \left( \frac{ R_{\rm in}/R_{\rm p} }{4} \right)^{-7/2} 
\end{eqnarray}
for the case that $T_{\rm p,e} \simeq T_{\rm int}$, 
where equation (\ref{eq:mdot_p_int}) is used, and 
\begin{eqnarray}
n_{\rm H}^{\rm therm}  & \simeq  & 
                                  4.7 \times 10^{13} {\rm cm}^{-3}\left( \frac{f_{\rm fill}}{10^{-2}} \right)^{-1}  \left( \frac{f_{\rm ram}}{1/ \sqrt{2}} \right)^{-3} \\ \nonumber
                         & \times &  \left( \frac{f_{\rm L}}{1/2} \right)^{2}   \left( \frac{1 - f_{\rm in}}{3/4} \right)^{2}  \left( \frac{ f_{\rm T} }{3/4} \right)^{-1/2} \\ \nonumber
                         & \times &   \left( \frac{M_{\rm p}}{10M_{\rm J}} \right)  \left( \frac{R_{\rm p}}{1.5R_{\rm J} } \right)^{-3} \left( \frac{ R_{\rm in}/R_{\rm p} }{4} \right)^{-21/2} \end{eqnarray}
for the case that $T_{\rm p,e} \simeq T_{\rm therm}$, 
where equation (\ref{eq:mdot_p_therm}) is used.

We now turn our attention to deriving a constrain on $R_{\rm in}$.
As discussed in Section \ref{sec:pla_mag},
disk limited gas accretion becomes energetic enough to self-regulate the thermal properties of other processes taking place in such a stage
when $T_{\rm p,e} \simeq T_{\rm therm}$.
For this case, conservation of energy dictates 
\begin{equation}
L_{\rm gas} \geq \frac{1}{2} \dot{M}_{\rm p} v_{\rm sh}^2,
\end{equation}
where the inequality sign appears as some energy may radiate away from magnetospheric flow (see Section \ref{sec:mod_en}).
The above equation is rewritten as (see equations (\ref{eq:f_in}) and (\ref{eq:gas_kin}))
\begin{equation}
 1- \frac{3}{2} \left( \frac{R_{\rm p}}{R_{\rm in}} \right) \left[1 - \frac{2}{3} \left( \frac{R_{\rm p}}{R_{\rm in}} \right)^{1/2} \right] \geq \left( 1 - \frac{R_{\rm p}}{R_{\rm in}} \right).
\end{equation}
This reads $R_{\rm in} \leq 4 R_{\rm p}$.
Note that the above condition is applicable only to the case that $T_{\rm p,e} \simeq T_{\rm therm}$;
for the case that $T_{\rm p,e} \simeq T_{\rm int}$, 
planetary magnetic fields and hence the disk truncation radius are determined by the intrinsic temperature 
(equations (\ref{eq:B_ps1}) and (\ref{eq:R_in_int}), also see Section \ref{sec:gas_flow}).
These quantities are controlled by earlier formation histories
and may provide additional heating for accretion flow.

In summary, magnetospheric flow around accreting giant planets can result in H$\alpha$ emission
either via accretion shock at planetary surfaces or via accretion flow that would be heated by inefficient cooling of certain lines.
Due to conservation of energy, $R_{\rm in} \leq 4 R_{\rm p}$ when disk-limited gas accretion self-regulates energetics of processes operating at such a stage.

\subsection{Gas flow from circumstellar disks} \label{sec:gas_flow}

We have so far focused on gas accretion flow in the vicinity of magnetized giant planets.
We here consider a more global configuration of accretion flow.
In particular, we explore what gas flow looks like from parental circumstellar disks onto accreting planets and the surrounding circumplanetary disks.
This consideration becomes important when planets and their circumplanetary disks are embedded in the circumstellar disks.
For this case, 
we can obtain another constraint on $R_{\rm in}$ that comes from the surrounding environment (e.g., stellar accretion rates).

Gas accretion flow from circumstellar disks onto circumplanetary disks and/or the host planets is poorly constrained currently.
The primary reason is that observations of circumplanetary disks are very limited to date;
disks around PDS 70 b/c are the only examples so far \citep[e.g.,][]{2019ApJ...879L..25I,2021ApJ...916L...2B}.
The lack of observations hinders specification of disk properties and hence development of reliable models.
Under this circumstance, we consider only an overall structure of gas accretion flow in this work.

We make use of the approach of \cite{2016ApJ...823...48T} to estimate how much of gas is delivered from parental circumstellar disks to the system of planets and their circumplanetary disks.
In the approach, the results of two different hydrodynamical simulations are coupled together;
one kinds of simulations derive a formula of the accretion rate onto such a system \citep{2002ApJ...580..506T},
and the other simulations compute a reduction factor of the surface density of circumstellar disks,
which is caused by disk-planet interaction \citep{2015ApJ...806L..15K}.
The resulting gas flow rate ($\dot{M}_{\rm p}^{\rm CSD} $) is given as
\begin{equation}
\label{eq:Mpdot_CSD}
\dot{M}_{\rm p}^{\rm CSD}  =           \frac{8.5}{3 \pi} \left( \frac{c_{\rm s}^{\rm CSD}}{v_{\rm Kep}^{\rm CSD}} \right) \left( \frac{M_{\rm p}}{M_{\rm s}} \right)^{-2/3}  \dot{M}_{\rm s}, 
\end{equation}
where $c_{\rm s}^{\rm CSD}$ and $v_{\rm Kep}^{\rm CSD}$ are the sound speed and the Keplerian velocity of the circumstellar disk gas at the position of planets, 
and $M_{\rm s}$ and $\dot{M}_{\rm s}$ are the mass of the central star and the disk accretion rate onto the star, respectively.

The value of $\dot{M}_{\rm p}^{\rm CSD} $ becomes comparable to the accretion rate onto planets ($\dot{M}_{\rm p}$),
if the accretion flow is in a steady state.
There is no guarantee that accreting giant planets achieve such a state.
However, \citet{2021ApJ...923...27H} have recently found that this might be the case for PDS 70 b/c;
their calculations show that 
\begin{eqnarray}
\label{eq:Mpdot_CSD2}
\dot{M}_{\rm p}^{\rm CSD}  & \simeq & 1.8 \times 10^{-7}  M_{\rm J} \mbox{ yr}^{-1} \left( \frac{c_{\rm s}^{\rm CSD} / v_{\rm Kep}^{\rm CSD}}{8.9 \times 10^{-2}} \right)   \\ \nonumber
                                         & \times  &   \left( \frac{M_{\rm p}}{10M_{\rm J}} \right)^{-2/3} \left( \frac{\dot{M}_{\rm s}}{1.4 \times 10^{-10} M_{\odot} \mbox{ yr}^{-1}}  \right) ,
\end{eqnarray}
where $c_{\rm s}^{\rm CSD}/v_{\rm Kep}^{\rm CSD}=8.9 \times 10^{-2}$ and $M_{\rm s}=0.76~ M_{\odot}$ are adopted, following \citet{2018A&A...617A..44K} 
which simulate the properties of the circumstellar disk around PDS 70,
and the value of $\dot{M}_{\rm s}$ is taken from \citet{2020ApJ...892...81T},
suggesting  that $\dot{M}_{\rm s}$ of PDS 70 lies within the range of $0.6-2.2 \times 10^{-10} M_{\odot} \mbox{ yr}^{-1}$.

It is noticeable that within the range of $\dot{M}_{\rm s}$,
the resulting value of $\dot{M}_{\rm p}^{\rm CSD}$ is comparable to the value of $\dot{M}_{\rm p} \simeq 10^{-8}-10^{-7} M_{\rm J} \mbox{ yr}^{-1}$,
which is estimated from the observed H$\alpha$ emission for PDS 70 b/c \citep[][references herein]{2021ApJ...923...27H}.
This implies that the steady state accretion assumption may not be unreasonable for accreting giant planets at least during certain formation stages.

Motivated by the finding, we use the assumption and derive a constraint on $R_{\rm in}$.
To proceed, we consider two limiting cases: $T_{\rm p,e} \simeq T_{\rm int}$ and $T_{\rm p,e} \simeq T_{\rm therm}$,
and equate $\dot{M}_{\rm p}^{\rm int}$ and $\dot{M}_{\rm p}^{\rm therm}$ with $\dot{M}_{\rm p}^{\rm CSD}$.
For the case that $T_{\rm p,e} \simeq T_{\rm int}$,
\begin{eqnarray}
\label{eq:R_in_int}
\frac{ R_{\rm in}^{\rm int} }{R_{\rm p}} & \simeq & 2.0 
                                                             \left( \frac{f_{\rm ram}}{1/ \sqrt{2}} \right)^{-2/7}  \left( \frac{h_{0} }{0.05 } \right)^{-2/7}   \\ \nonumber
                                       & \times   &  \left( \frac{M_{\rm p}}{10 M_{\rm J} } \right)^{1/7}  \left( \frac{R_{\rm p}}{1.5R_{\rm J} } \right)^{3/7} \left( \frac{T_{\rm int}}{700 {\rm K}} \right)^{16/21}    
                                                           \\ \nonumber
                                        & \times   &  \left( \frac{M_{\rm s} }{1 M_{\odot} } \right)^{-4/21}   \left( \frac{ \dot{M}_{\rm s} }{ 10^{-10} M_{\odot} \mbox{ yr}^{-1}} \right)^{-2/7}     \\ \nonumber 
                                          & \times   &    \left( \frac{r_{\rm p} }{50 \mbox{ au} } \right)^{-1/14},
\end{eqnarray}
where equations (\ref{eq:mdot_p_int}) and (\ref{eq:Mpdot_CSD}) are used, and 
for the case that $T_{\rm p,e} \simeq T_{\rm therm}$,
\begin{eqnarray}
\label{eq:R_in_therm}
\frac{ R_{\rm in}^{\rm therm} }{R_{\rm p}}& \simeq & 4.0   \left( \frac{f_{\rm ram}}{1/ \sqrt{2}} \right)^{-2/7} \left( \frac{f_{\rm L}}{1/2} \right)^{4/21}
                                                \left( \frac{h_{0} }{0.05 } \right)^{-2/21}  \\ \nonumber
                         & \times &     \left( \frac{1 - f_{\rm in}}{3/4} \right)^{4/21} \left( \frac{M_{\rm p}}{10M_{\rm J} } \right)^{13/63}  \left( \frac{R_{\rm p}}{1.5R_{\rm J} } \right)^{-1/7} \\ \nonumber
                         & \times &    \left( \frac{M_{\rm s}}{1M_{\odot}} \right)^{-4/63}  \left( \frac{ \dot{M}_{\rm s} }{ 10^{-10} M_{\odot} \mbox{ yr}^{-1}} \right)^{-2/21}  \\ \nonumber
                           & \times   &     \left( \frac{r_{\rm p} }{50 \mbox{ au} } \right)^{-1/42},
\end{eqnarray}
where equations (\ref{eq:mdot_p_therm}) and (\ref{eq:Mpdot_CSD}) are used.
Note that in the above equations, $c_{\rm s}^{\rm CSD}/v_{\rm Kep}^{\rm CSD} \equiv h_{0} (r_{\rm p}/ 1 \mbox{ au})^{1/4}$ is assumed,
where $h_0=0.05$, and $r_{\rm p}$ is the position of planets.
Also, the dependence on $1 - f_{\rm in}$ is weak, and an intermediate value of $3/4$ is used in equation (\ref{eq:R_in_therm});
the value of $1 - f_{\rm in}$ varies from 1/2 to 1.

\begin{figure}
\begin{center}
\includegraphics[width=8.3cm]{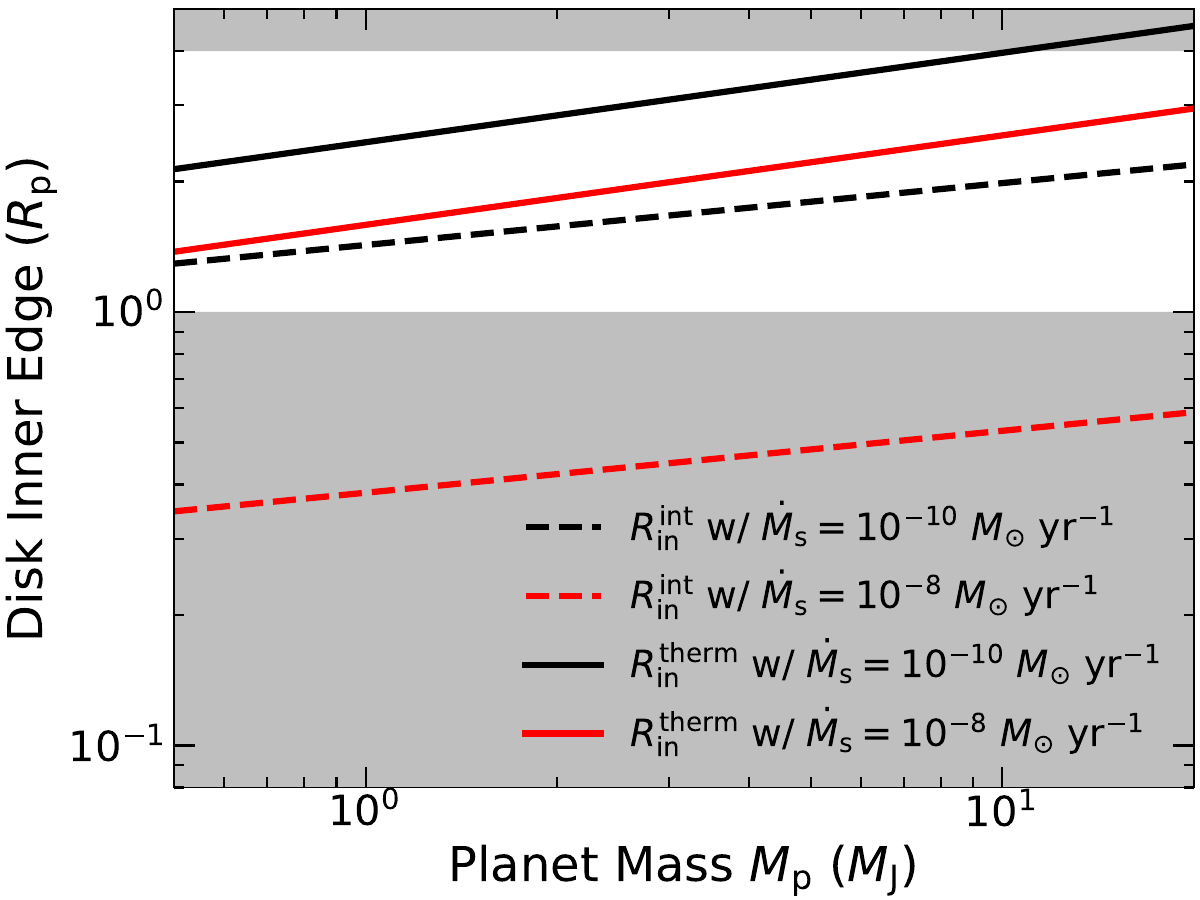}
\caption{The computed value of $R_{\rm in}$ as a function of $M_{\rm p}$ for a given value of $\dot{M}_{\rm s}$.
As examples, $\dot{M}_{\rm s}=10^{-8} M_{\odot}$ yr$^{-1}$ and $\dot{M}_{\rm s}=10^{-10} M_{\odot}$ yr$^{-1}$ are picked.
The prohibited regions (i.e., $R_{\rm in} \leq 1 R_{\rm p}$ and $R_{\rm in} > 4 R_{\rm p}$) are denoted by the grey shaded regions.
High stellar accretion rates shrink $R_{\rm in}$, while it expands for massive planets.
Magnetospheric accretion is viable for a wide range of parameters in the case that  $T_{\rm p,e} \simeq T_{\rm therm}$, 
as $1 < R_{\rm in}^{\rm therm} \leq 4$ when $M_{\rm p} \leq 10 M_{\rm J}$.
In contrast, it becomes possible only at the later stages of disk evolution for the case that  $T_{\rm p,e} \simeq T_{\rm int}$.}
\label{fig3}
\end{center}
\end{figure}

Figure \ref{fig3} visualizes how $R_{\rm in}^{\rm int}$ and $R_{\rm in}^{\rm therm}$ change as a function of $M_{\rm p}$ for a given value of $\dot{M}_{\rm s}$;
since the dependence on other parameters including $r_{\rm p}$ is very weak (see equations (\ref{eq:R_in_int}) and (\ref{eq:R_in_therm})), 
we focus on $M_{\rm p}$ and $\dot{M}_{\rm s}$.
Also, $1 - f_{\rm in}$ is set at $3/4$ as done in equation (\ref{eq:R_in_therm}).
The resulting trends can be understood readily;
when stellar accretion rates are high, the ram pressure becomes strong, and hence the disk inner edge locates close to the host planets.
For the $M_{\rm p}$ dependence, monotonic increase of $R_{\rm in}$ arises due to disk-planet interaction (equation (\ref{eq:Mpdot_CSD}));
massive planets open up a deep gap in their parental circumstellar disks, which decreases the gas flow ($\dot{M}_{\rm p}^{\rm CDS}$) 
onto these planets and the surrounding circumplanetary disks.
As a result, $R_{\rm in}$ expands due to low ram pressure.
Our calculations show that when $T_{\rm p,e} \simeq T_{\rm therm}$, $1 < R_{\rm in}^{\rm therm} \leq 4$ for planets with $M_{\rm p} \leq 10 M_{\rm J}$,
suggesting that magnetospheric accretion is possible for a wide range of parameters.
On the other hand, when $T_{\rm p,e} \simeq T_{\rm int}$, $R_{\rm in}^{\rm int} < 1$ for high stellar accretion rates.
Therefore, magnetospheric accretion occurs only in the later stage of disk evolution.

The condition ($R_{\rm in}^{\rm int} / R_{\rm p} > 1$) needed for the case that $T_{\rm p,e} \simeq T_{\rm int}$ leads to a constraint on $\dot{M}_{\rm s}$ as
\begin{eqnarray}
\label{eq:Mdot_s_int}
\dot{M}_{\rm s} & < & 1.1 \times 10^{-9} M_{\odot} \mbox{ yr}^{-1} 
                                                                           \left( \frac{f_{\rm ram}}{1/ \sqrt{2}} \right)^{-1}  \left( \frac{ h_0 }{ 0.05 } \right)^{-1}   \\ \nonumber
                                       & \times   & \left( \frac{M_{\rm p}}{10M_{\rm J}} \right)^{1/2}  \left( \frac{R_{\rm p}}{1.5R_{\rm J} } \right)^{3/2} \left( \frac{T_{\rm int}}{700 {\rm K}} \right)^{8/3} \\ \nonumber
                                       & \times  &  \left( \frac{M_{\rm s}}{1 M_{\odot}} \right)^{-2/3}  \left( \frac{ r_{\rm p} }{50 \mbox{ au}} \right)^{-1/4},
\end{eqnarray}
where equations (\ref{eq:mdot_p_int}) and (\ref{eq:Mpdot_CSD}) are used.
This is equivalent to the condition that $\dot{M}_{\rm p}^{\rm int} > \dot{M}_{\rm p}^{\rm CSD}$ at $R_{\rm in} = R_{\rm p}$.
For the case that $T_{\rm p,e} \simeq T_{\rm therm}$, the required condition ($R_{\rm in}^{\rm therm} / R_{\rm p} \leq 4$) is rewritten as
\begin{eqnarray}
\label{eq:Mdot_s_therm}
\dot{M}_{\rm s} & \geq & 9.0 \times 10^{-11} M_{\odot} \mbox{ yr}^{-1} \left( \frac{f_{\rm ram}}{1/ \sqrt{2}} \right)^{-3} \left( \frac{f_{\rm L}}{1/2} \right)^{2}   \\ \nonumber
                                            & \times &    \left( \frac{ h_0 }{ 0.05 } \right)^{-1}    \left( \frac{1 - f_{\rm in}}{3/4} \right)^{2}    \left( \frac{M_{\rm p}}{10M_{\rm J}} \right)^{13/6}  \left( \frac{R_{\rm p}}{1.5R_{\rm J} } \right)^{-3/2}  \\ \nonumber
                                       & \times  &  \left( \frac{M_{\rm s}}{1 M_{\odot}} \right)^{-2/3}  \left( \frac{ r_{\rm p} }{50 \mbox{ au}} \right)^{-1/4},
\end{eqnarray}
where equations (\ref{eq:mdot_p_therm}) and (\ref{eq:Mpdot_CSD}) are used.
This can also be obtained from the condition that $\dot{M}_{\rm p}^{\rm therm} \leq \dot{M}_{\rm p}^{\rm CSD}$ at $R_{\rm in} = 4 R_{\rm p}$.

In the following section, we use equations (\ref{eq:Mdot_s_int}) and (\ref{eq:Mdot_s_therm}) 
and predict when hydrogen lines can be emitted from young giant planets via magnetospheric accretion,
either due to accretion shock or the inefficiently cooled accretion flow.

\subsection{Predicted line luminosity} \label{sec:line_lum}

Armed with equations derived in the above sections,
we are now ready to explore the line luminosity of hydrogen emission originating from giant planets undergoing magnetospheric accretion.
In order to compute the line luminosity, we heavily rely on relationships between line and accretion luminosities that are obtained by previous studies:
\citet{2021ApJ...917L..30A} for emission from accretion shock and \citet{2017A&A...600A..20A} for emission from accretion flow along magnetospheres.
The former computes the relationship theoretically, and the latter obtains it observationally from CTTSs.

We first summarize the key equations to compute the accretion luminosity.
Since we target planets surrounded by their circumplanetary disks, which are embedded in their parental circumstellar disks,
we assume that  $\dot{M_{\rm p}} \simeq \dot{M}_{\rm p}^{\rm CSD}$ as discussed in Section \ref{sec:gas_flow}.
Then, the accretion luminosity is written as
\begin{eqnarray}
\label{eq:Lacc}
 L_{\rm acc} & \simeq & 5.2 \times 10^{-4} L_{\odot} \left( \frac{h_{0} }{0.05 } \right) \left( \frac{M_{\rm p}}{10 M_{\rm J} } \right)^{1/3}  \left( \frac{R_{\rm p}}{1.5R_{\rm J} } \right)^{-1}   \\ \nonumber
                     & \times   &                      \left( \frac{M_{\rm s} }{1 M_{\odot} } \right)^{2/3}   \left( \frac{ \dot{M}_{\rm s} }{ 10^{-10} M_{\odot} \mbox{ yr}^{-1}} \right)     
                                                              \left( \frac{r_{\rm p} }{50 \mbox{ au} } \right)^{1/4},
\end{eqnarray}
where equations (\ref{eq:L_acc}) and (\ref{eq:Mpdot_CSD}) are used.

We then consider two limiting cases for $T_{\rm p,e}$:
when planets undergo magnetospheric accretion and their effective temperature is given as $T_{\rm p,e} \simeq T_{\rm int}$,
the accretion luminosity can be emitted only when $1 < R_{\rm in}^{\rm int}/R_{\rm p} $ (see equations (\ref{eq:R_in_int}) and (\ref{eq:Mdot_s_int})).
Magnetic fields of these planets are written by equation (\ref{eq:B_ps1}).
On the other hand, when planets with the effective temperature of $T_{\rm p,e} \simeq T_{\rm therm}$ (see equation (\ref{eq:Tpe_therm})) experience magnetospheric accretion,
then the accretion luminosity can be emitted only when $ (1 < ) R_{\rm in}^{\rm therm}/R_{\rm p} \leq 4$ (see equations (\ref{eq:R_in_therm}) and (\ref{eq:Mdot_s_therm})).
Their magnetic fields are given by equation (\ref{eq:Bps_therm}).

\begin{figure}
\begin{center}
\includegraphics[width=8.3cm]{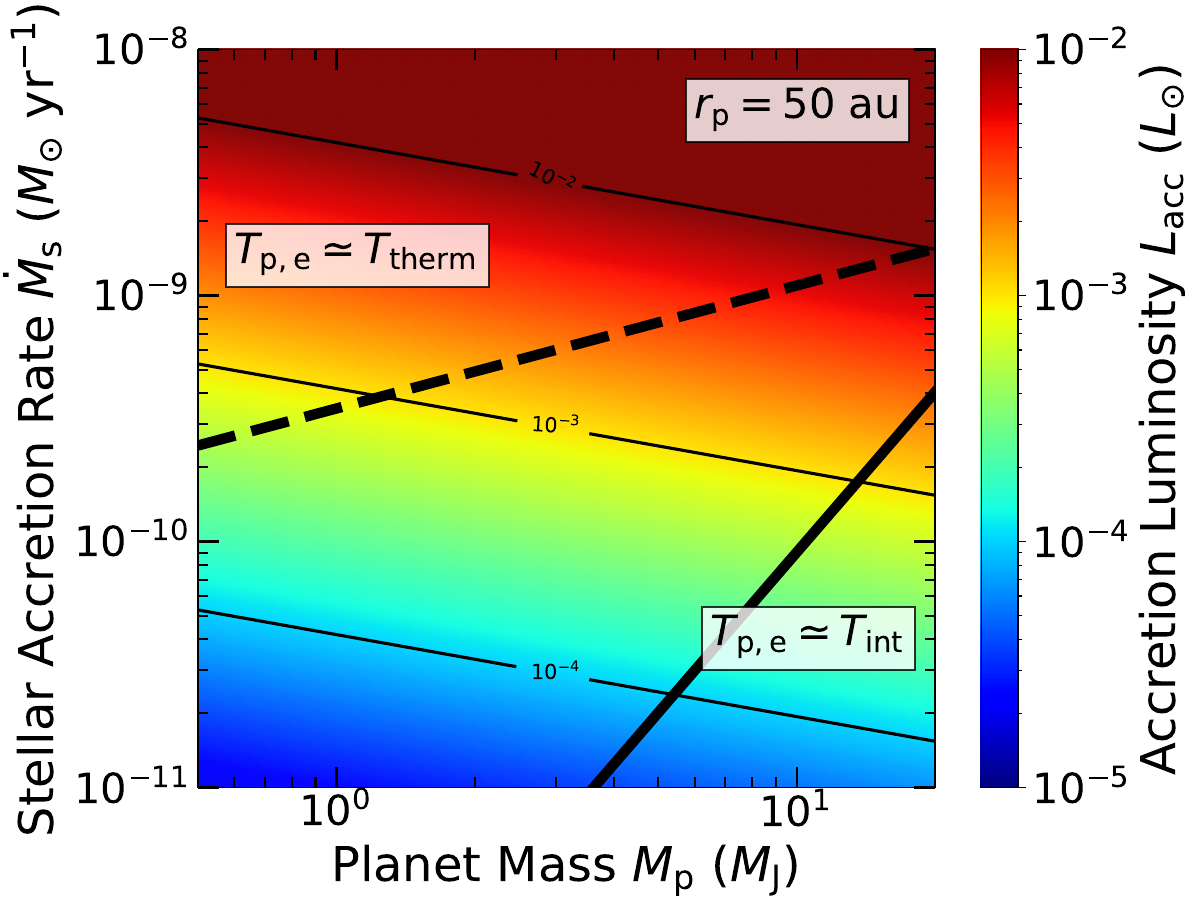}
\caption{The computed value of $L_{\rm acc}$ as a function of $M_{\rm p}$ and $\dot{M}_{\rm s}$.
As an example, $r_{\rm p}= 50$ au is chosen.
The observability of $L_{\rm acc}$ increases for high planet masses and high stellar accretion rates.
The ultimate origin of $L_{\rm acc}$ can be identified if observed systems are located in 
either the region above the dashed line (see equation (\ref{eq:Mdot_s_int})) or the region below the solid line (see equation (\ref{eq:Mdot_s_therm})).}
\label{fig4}
\end{center}
\end{figure}

Separating the contributions of $T_{\rm p,e}$ ($T_{\rm int}$ vs $T_{\rm therm}$ in equation (\ref{eq:T_pe})) allows us to identify a parameter space, 
wherein each contribution becomes dominant.
This is clearly shown in Figure \ref{fig4} that plots under what conditions, what value of $L_{\rm acc}$ can be emitted from giant planets via magnetospheric accretion.
The value of $L_{\rm acc}$ increases monotonically with increasing $M_{\rm p}$ and $\dot{M}_{\rm s}$, which is obvious from equation (\ref{eq:Lacc}).
As anticipated from the above discussion, the $M_{\rm p}-\dot{M}_{\rm s}$ parameter space is divided into three regions:
the region above the dashed line, where disk-limited gas accretion is energetic enough to self-regulate the resulting line emission (i.e., $T_{\rm p,e} \simeq T_{\rm therm}$),
the region below the solid line, where early formation histories play an important role for $L_{\rm acc}$ even at the disk-limited gas accretion stage (i.e., $T_{\rm p,e} \simeq T_{\rm int}$),
and the intermediate region, where  both the cases are possible.
Our calculations show that magnetospheric accretion leads to $L_{\rm acc}$ that can be high enough to be observed for certain combinations of parameters 
(i.e., high $M_{\rm p}$ and high $\dot{M}_{\rm s}$).

\begin{table}
\begin{center}
\caption{Relationships between line and accretion luminosities}
\label{table2}
{
\begin{tabular}{c|c|ccc|ccc}
\hline
                       &                                      & \multicolumn{3}{c|}{Accretion shock}          & \multicolumn{3}{c}{Accretion flow}      \\ \hline 
Line                &  $\lambda$ ($\mu$m)   & $a$       & $b$    & $b/a$                             &  $a$       & $b$    & $b/a$                          \\  \hline
H$\alpha$      &  0.656                            & 0.95     & 1.61    & 1.69                               &  1.13      &1.74    & 1.54                           \\           
Pa$\beta$      &  1.282                            & 0.86     & 2.21    & 2.57                               &  1.06      & 2.76   & 2.60                         \\           
Br$\gamma$  &  2.166                            & 0.85     & 2.84    & 3.34                               &  1.19      & 4.02   & 3.38                            \\                              
\hline                                 
\end{tabular}

The values of $a$ and $b$ are adopted from \citet{2021ApJ...917L..30A} and \citet{2017A&A...600A..20A} for accretion shock and flow, respectively.
}
\end{center}
\end{table}

We now compute the line luminosity ($L_{\rm line}$) of hydrogen emission, using the following equation:
\begin{equation}
\label{eq:L_line}
\log_{10} (L_{\rm acc}/ L_{\odot}) = a \times \log_{10} (L_{\rm line}/L_{\odot}) +b,
\end{equation}
where fitting parameters ($a$ and $b$) are summarized in Table \ref{table2}.
As an example, we consider three lines that tend to be observed readily (see Table \ref{table2}).
Similar calculations are straightforward for other lines \citep[see table 1 of][where the values of $a$ and $b$ are tabulated for other lines]{2021ApJ...917L..30A}.

\begin{figure*}
\begin{minipage}{17cm}
\begin{center}
\includegraphics[width=8.3cm]{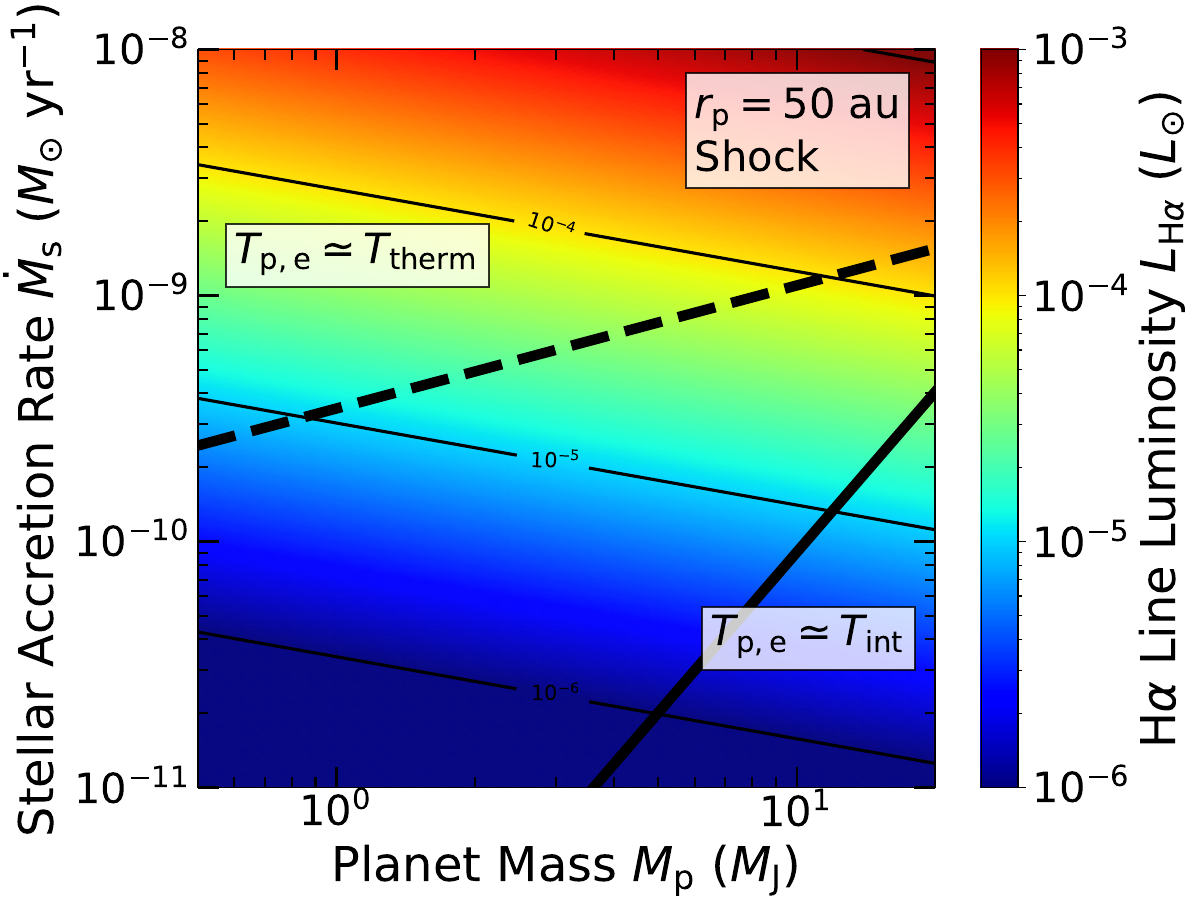}
\includegraphics[width=8.3cm]{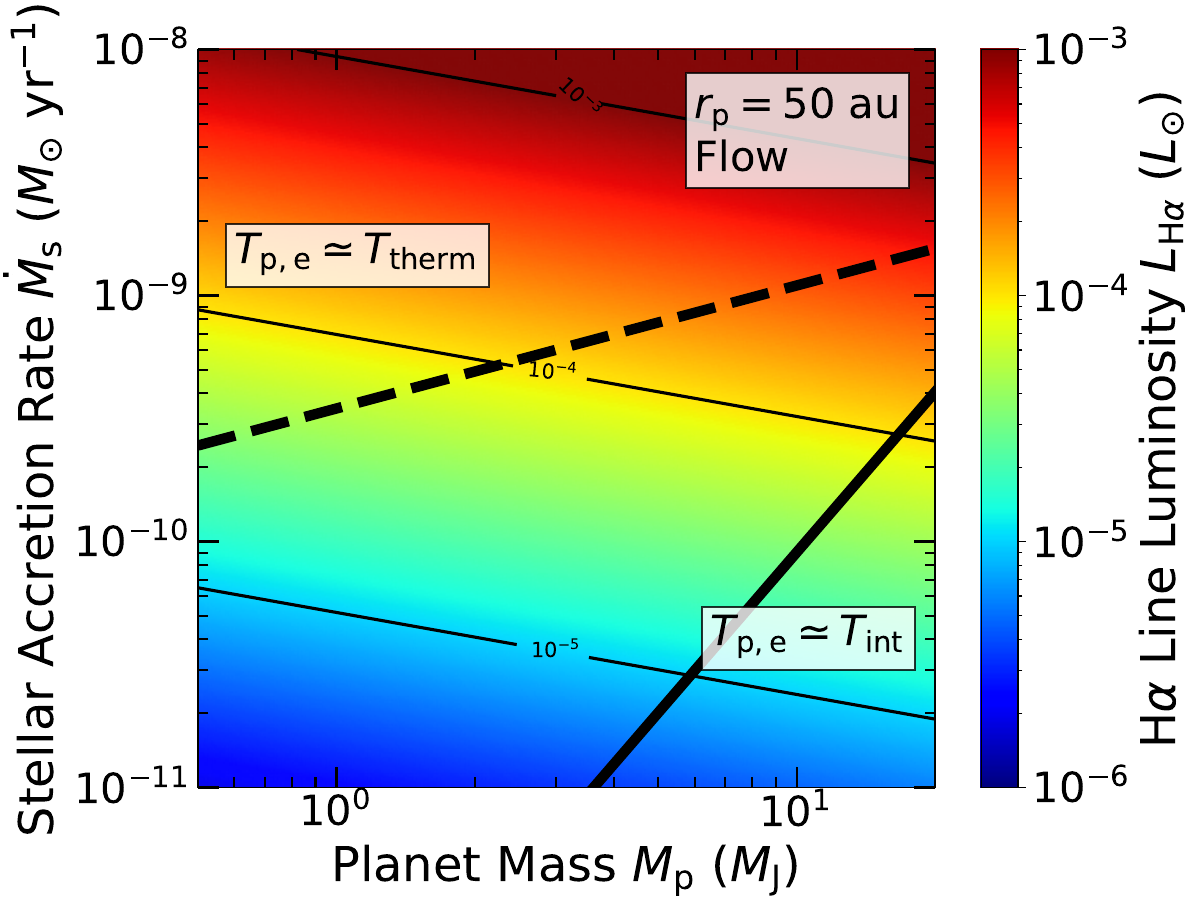}
\includegraphics[width=8.3cm]{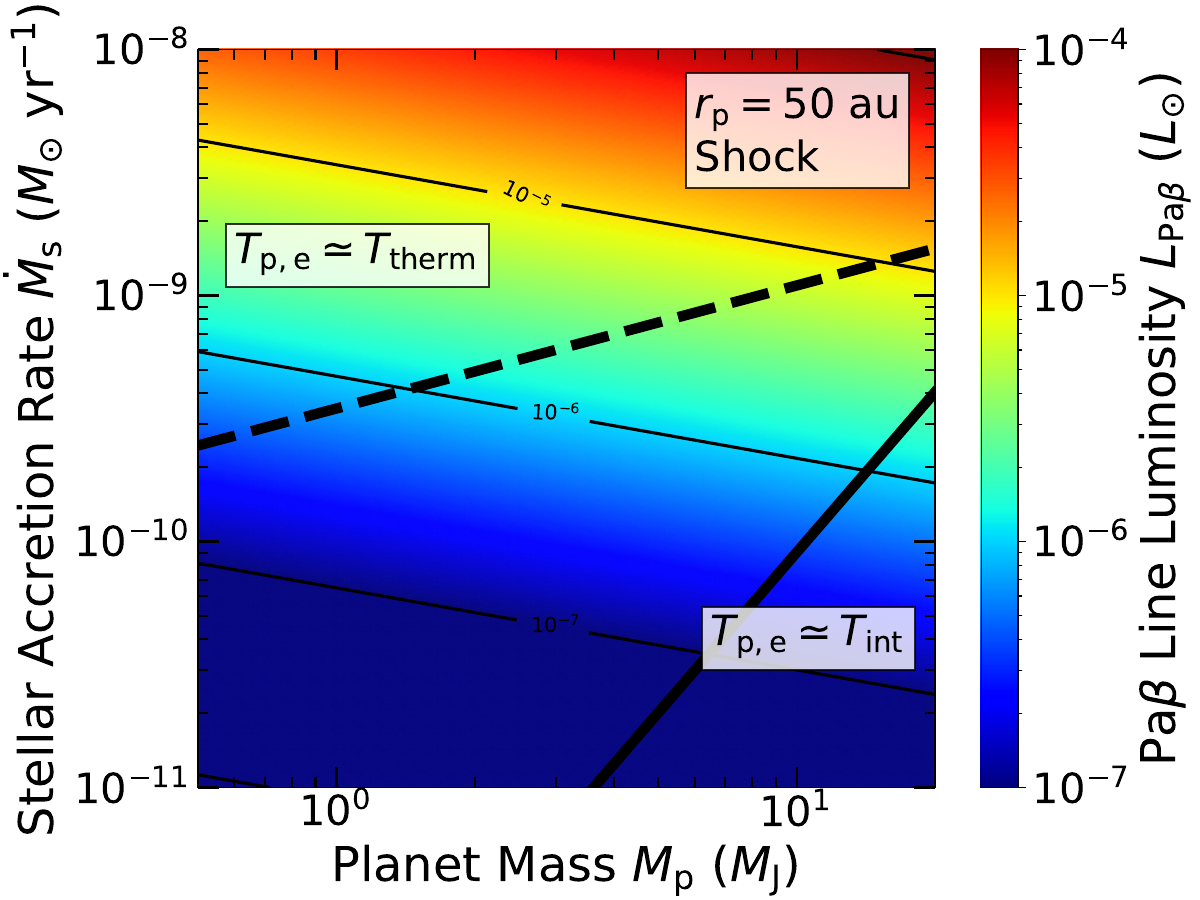}
\includegraphics[width=8.3cm]{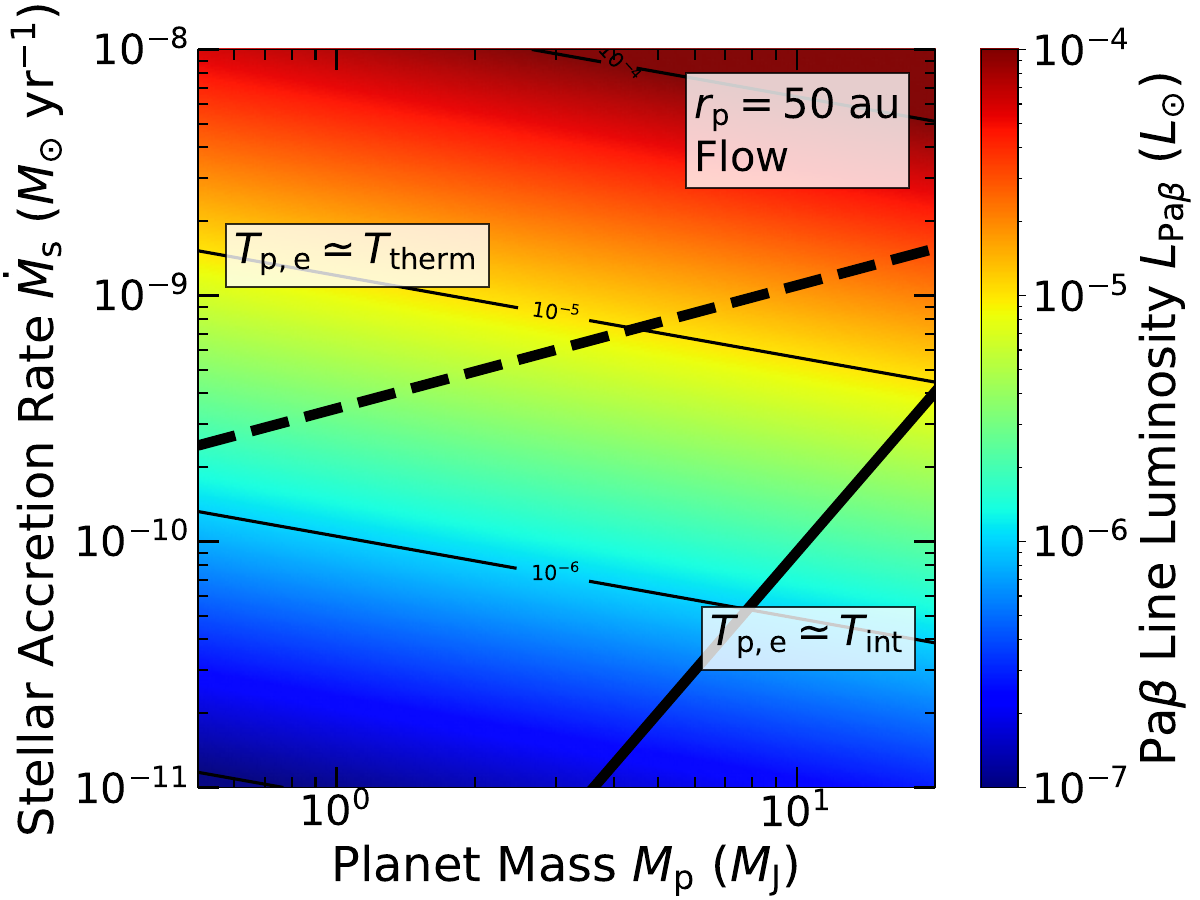}
\includegraphics[width=8.3cm]{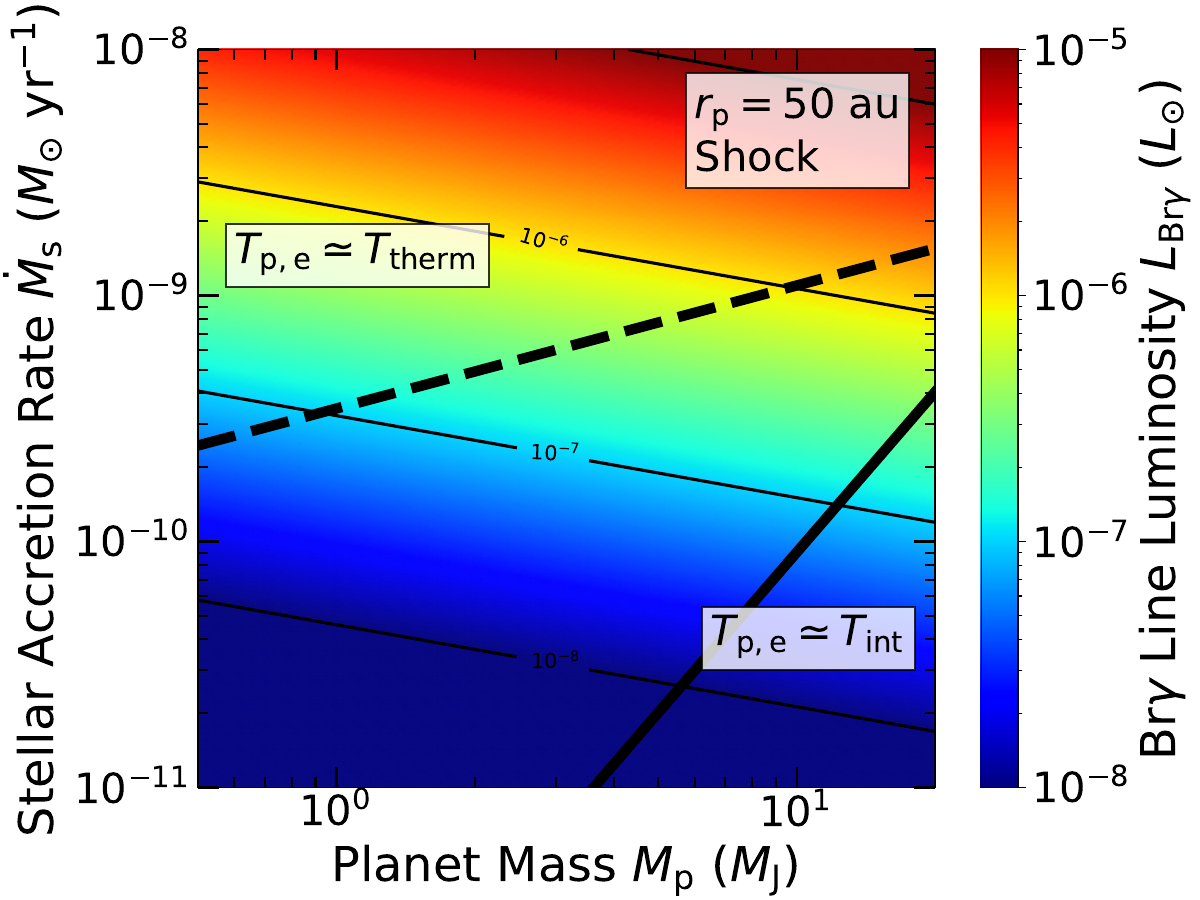}
\includegraphics[width=8.3cm]{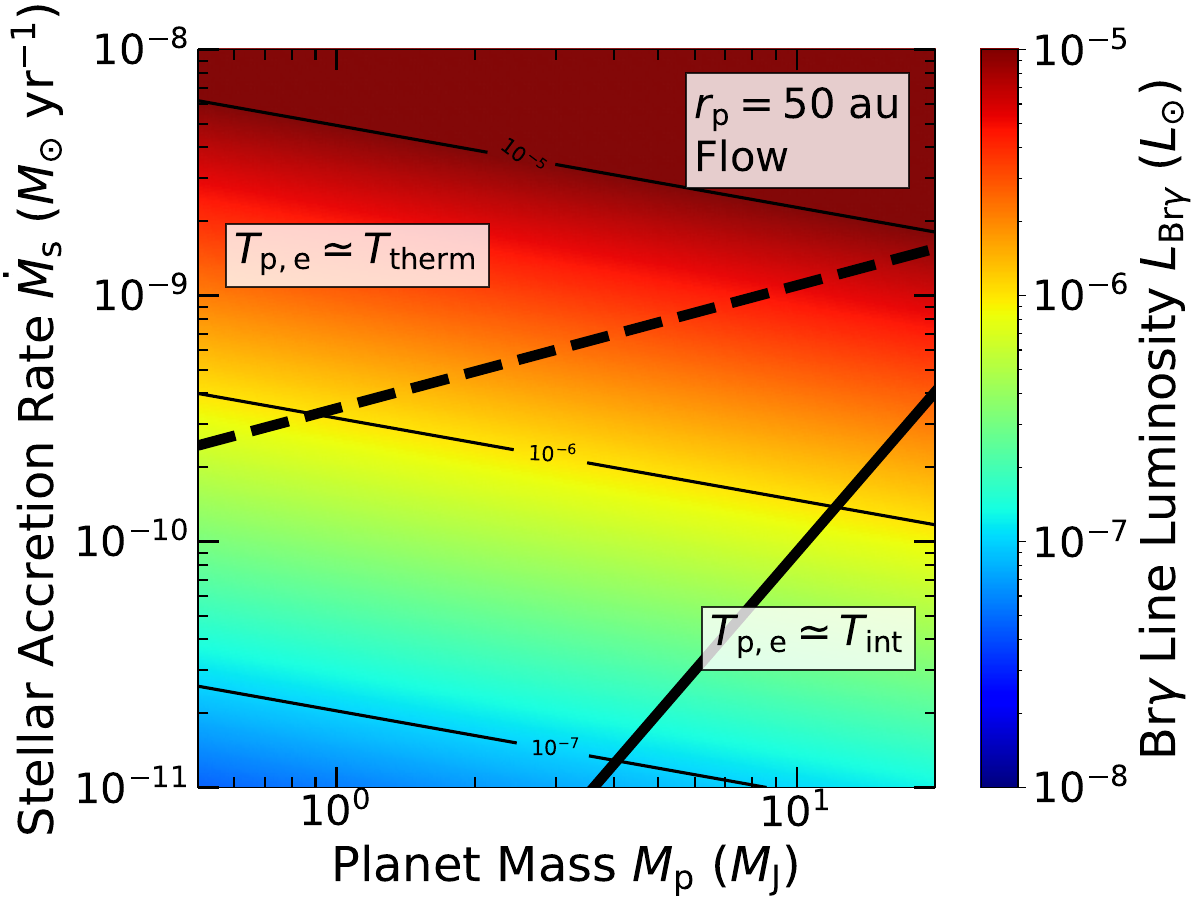}
\caption{The resulting line luminosities of accreting giant planets due to magnetospheric accretion.
As done in Figure \ref{fig4}, the planet position is set at $r_{\rm p}=50$ au.
On the left, the luminosities coming from accretion shock are plotted, while on the right, the ones originating from accretion flow are depicted.
From the top to the bottom, H$\alpha$, Pa$\beta$, and Br$\gamma$ luminosities are shown.
As expected, the accretion flow case leads to higher luminosities than the accretion shock case (see Table \ref{table2}).
Also, the line luminosities become weaker from the top to the bottom panels.
These line luminosities (and line ratios) become a theoretical prediction of when and how accreting giant planets emit (observable) hydrogen emission lines.}
\label{fig5}
\end{center}
\end{minipage}
\end{figure*}

Figure \ref{fig5} shows what the resulting line luminosities look like in the $M_{\rm p}-\dot{M}_{\rm s}$ parameter space.
The value of $r_{\rm p}=50$ au is picked as in Figure \ref{fig4}.
It is obvious from Table \ref{table2} that H${\alpha}$ line luminosity is more than one order magnitude dimmer than the accretion luminosity,
and the line luminosity is higher for the accretion flow case than the accretion shock case.
For Pa$\beta$ and Br$\gamma$, similar trends are confirmed, 
while $L_{{\rm Pa}\beta}$ and $L_{{\rm Br}\gamma}$ are more than two order and three order magnitudes lower than $L_{\rm acc}$, respectively.

In summary, one can predict what line luminosities of hydrogen will be emitted from young giant planets due to magnetospheric accretion,
if the accretion rate onto the host star and the mass and position of planets are estimated.
The observed value of line luminosities (and line ratios) can be used as a diagnostics to identify where the emission originates (planetary surface vs accretion flow)
and how the emission is produced (accretion shock vs accretion heating).
Specification of stellar accretion rates and planet properties enables determination of the ultimate origin of why such planets undergo magnetospheric accretion,
namely, the magnetism of young, accreting giant planets;
disk-limited gas accretion is energetic enough to trigger it, or early planet formation processes keep planets hot enough.

\section{Observational test} \label{sec:data}

We here apply our predictions made in Section \ref{sec:mod} to actual systems that can be observed.
To proceed, we conduct new observations targeting HD~163296 with Subaru/SCExAO+VAMPIRES.

\subsection{Observations and data reduction}

We observed HD~163296 on 2021 May 8 UT with Subaru/SCExAO+VAMPIRES under the NASA-Keck time exchange program (PID 61/2021A\_N200: PI - Hasegawa).

VAMPIRES has two detectors that can take different images with two filters simultaneously, 
and is capable of mitigating aberrations between the detectors by switching the filters \citep[double-differential calibration;][]{Norris2015}. 
When conducting H$\alpha$ imaging with VAMPIRES,
we used narrow-band filters for H$\alpha$ ($\lambda_{\rm c}=656.3$ nm, $\Delta\lambda=1.0$~nm) and adjacent continuum ($\lambda_{\rm c}=647.68$~nm, $\Delta\lambda=2.05$~nm), 
which allows us to effectively subtract continuum components from the H$\alpha$ image \citep[spectral differential imaging (SDI);][]{Smith1987}.
During the observations, we repeated two states for double-differential calibration;
in State~1, cam1 is used for continuum, and cam2 is for H$\alpha$,
and in State~2, the setup is the other way around. 
Note that due to an instrumental constraint,
we used a smaller field of view (FoV: $\sim1\farcs5\times1\farcs5$) than the maximum FoV of VAMPIRES ($\sim3\arcsec\times3\arcsec$).

The single exposure time was 50-msec in the first sequence (9 cubes in both State~1 and State~2), 
and from the second sequence, we changed it to 40~msec in order to avoid saturation. 
The difference of the single exposure time was corrected before post-processing. 
The total integration time corresponds to 2806~sec and 2369~sec, and the field rotation angle gains $\sim64^\circ$ and 62$^\circ$ for States~1 and 2, respectively.

The VAMPIRES data format is a cube consisting of an image and short exposures (2001 exposures per cube). 
We first subtracted dark from each exposure and conducted point spread function (PSF) fitting of continuum frames by 2D-Gaussian for frame selection. 
The typical full width at half maximum (FWHM) of the PSF was measured at $\sim20-25$~mas with a pixel scale of 6.24$\pm$0.01~mas/pix \citep{2022NatAs...6..751C}.
We investigated the fitted peaks and removed a few data cubes that do not exhibit the typical peaks due to poor-AO corrections. 
We then empirically selected 80 percentile of the fitted peak values in a cube and then combined the selected exposures into an image 
after aligning the centroid of the PSFs \citep[see Figure 2 of][]{Uyama2020}.

As for post-processing to remove stellar halo and to search for faint accretion signatures, 
we followed post-processing methods of \citet[][see their section 2.1 for details, references herein]{Uyama2022},
utilizing angular differential imaging \citep[ADI;][]{Marois2008}, SDI with the two filters, and the VAMPIRES double-differential calibration techniques. 
We scaled the continuum images, by calculating a scaling factor from comparison between the photometry (aperture radius = 10FWHM) of the H$\alpha$ and continuum filters
and by correcting wavelengths so that they can be appropriate for the reference PSF of the SDI reduction.
In the ADI reduction, we used {\tt pyklip} packages \citep[][]{pyklip} that make a reference PSF by Karhunen-Lo\`eve Image Projection \citep[KLIP;][]{Soummer2012}.
Note that we adopted aggressive ADI reduction to explore as faint accretion signatures as possible, 
and this setting can attenuate extended features. 
Therefore, we do not discuss the H$\alpha$ jet features that are present in \cite{Xie2021}. 
We also note that the A4 knot detected by \cite{Xie2021} is out of the FoV in our observations.
We then applied SDI using the ADI residuals of H$\alpha$ and scaled-continuum at both State~1 and 2, and finally conducted double differential calibration.

\subsection{Observational results}

\begin{figure}
\begin{center}
\includegraphics[width=10cm]{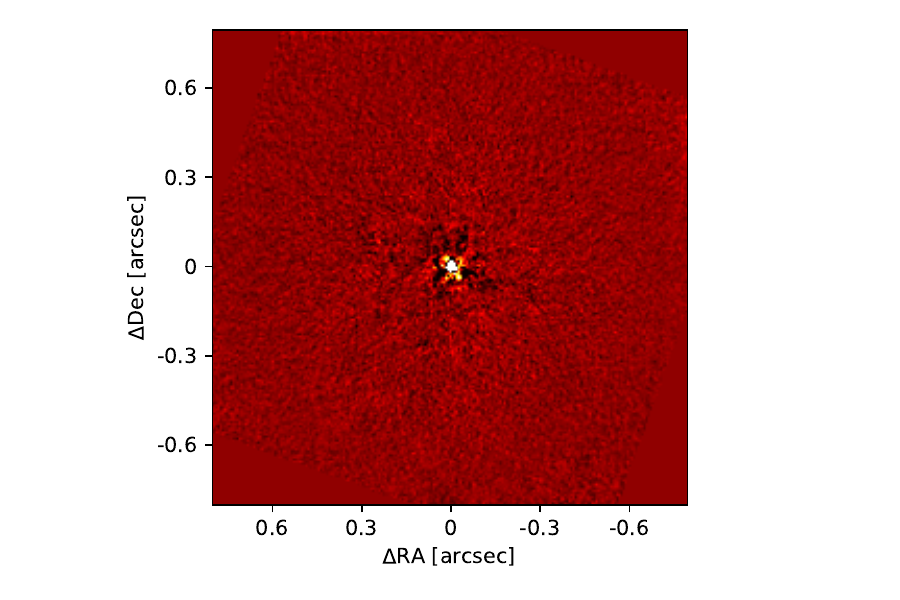}
\caption{The post-processed H$\alpha$ image of HD~163296 taken by the Subaru/VAMPIRES.
The value of KL=20 in {\tt pyklip}-ADI reduction is used. 
In the image, the north is up and the east is left. 
The central star is masked by the algorithm.
No point-like sources emitting H$\alpha$ are discovered.}
\label{fig6}
\end{center}
\end{figure}

\begin{figure}
\begin{center}
\includegraphics[width=8.3cm]{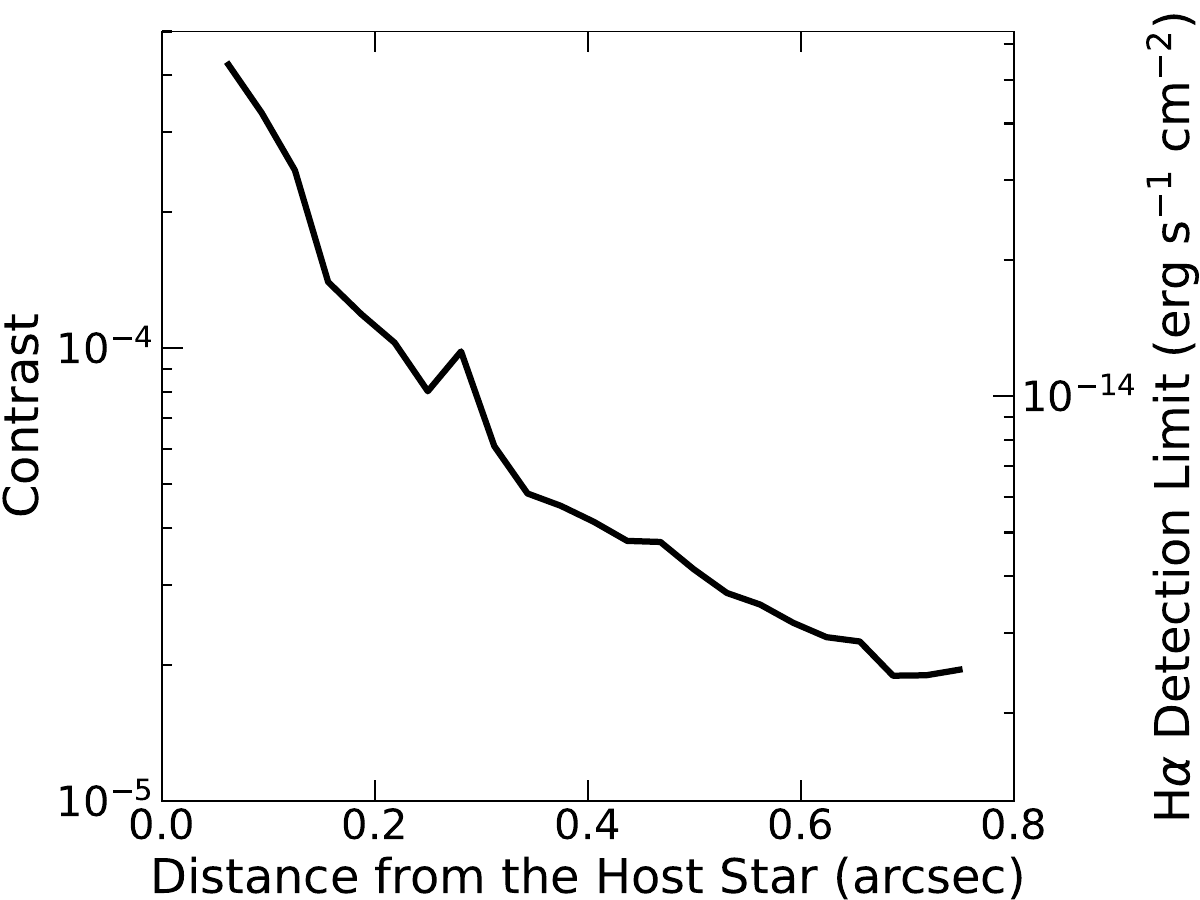}
\caption{The contrast and the corresponding 5$\sigma$ detection limit of H$\alpha$ line 
as a function of the distance from the host star for our observations on the left and right axes, respectively.}
\label{fig7}
\end{center}
\end{figure}

We did not find any companion candidates within 0\farcs7.\footnote{
Recently, the presence of faint H$\alpha$ emission is reported \citep{2022A&A...668A.138H}.
Its origin is unclear and our observations achieved a better detection limit (see Figure \ref{fig7} vs their figure 6).}
This is clearly shown in Figure \ref{fig6}.
We calculated standard deviations within annular regions after convolving the output image with a radius of FWHM/2.
The image is then compared with photometry of the central star with the aperture radius of FWHM/2 at the H$\alpha$ filter for a contrast limit (Figure \ref{fig7}).
We also took into account throughput loss made by the ADI reduction, where fake PSFs are injected.

In order to compare the observation results with theoretical predictions directly,
we convert the contrast limit into the H$\alpha$ flux limit.
The conversion is done by referring to the continuum flux of the central star and taking into account a H$\alpha$/continuum ratio from our observations. 
The resulting conversion factor is $\sim 1.3 \times 10^{-10}$.
Note that the detection limit corresponds to the integrated line flux of H$\alpha$ as our observations cannot resolve the line, 
and thus we do not take into account the line profile for the comparison between the observational results and our model. 
The VAMPIRES H$\alpha$ narrow-band filter with 1.0~nm corresponds to a velocity coverage of $\pm100$~km/s, 
which is well above the possible maximum gas velocity around a Jovian protoplanet \citep[see also Appendix A in][]{Uyama2020}.
\cite{Sitko2008} present that HD~163296 exhibits no variability within 10~\%, 
except for a significant variability in the NIR wavelengths per 16~years. 
We therefore used {\it Gaia} G-band flux \citep[4.81 erg~s$^{-1}$~cm$^{-2}$~\micron$^{-1}$;][]{Gaia-DR2} as the continuum. 
The H$\alpha$/continuum ratio is estimated at 2.66 
by comparing the photometry of the central star between the H$\alpha$ and the continuum filters, 
and assuming the multiplied value as the HD~163296 flux at H$\alpha$. 
The aperture radius of 10FWHM is used in the above conversion.

Figure \ref{fig7} shows the 5$\sigma$ detection limit as a function of the distance from the host star. 
Note that the detection limit is not computed by \cite{Xie2020,Xie2021},
where the MUSE data taken toward HD~163296 are analyzed;
those data include instrumental noises, which makes it hard to accurately estimate the detection limit. 
Direct comparison between our detection limit and the MUSE one is thus not made in this work.

In the following sections, we use the above detection limit and apply our theoretical predictions to the HD~163296 system.

\subsection{Effects of extinction} \label{sec:extinction}

Before comparing our predictions with the observational results,
we here consider the effect of extinction.

Extinction occurs when gas and/or dust are present between emitting sources and the observer,
which can potentially reduce the observed line flux significantly from the intrinsically emitted flux.
Its value ($A_{\lambda}$) measured in magnitude at a wavelength $\lambda$ is defined as \citep[e.g.,][]{2011piim.book.....D}
\begin{equation}
\label{eq:ext_A1}
A_{\lambda} \equiv 2.5 \log_{10} \left( F_{\lambda} / F_{\lambda}^{\rm obs} \right),
\end{equation}
where $F_{\lambda}^{\rm obs}$ is the actually observed flux, and $F_{\lambda}$ is the intrinsic flux emitted from the sources before extinction comes into play.
In this work, $F_{\lambda}$ corresponds to the theoretically computed value.

The value of $A_{\lambda}$ is quantified relatively well for star-forming environments \citep[e.g.,][]{2011piim.book.....D}.
In fact, $A_{\lambda}$ is written as 
\begin{equation}
\label{eq:ext_A2}
A_{\lambda}  = N_{\rm H} / K_{\lambda},
\end{equation}
where $N_{\rm H}$ is the total column density of hydrogen distributing between the sources and the observer,
and $K_{\lambda}$ is the conversion coefficient.
For diffuse ISM (interstella medium) and molecular clouds,
the value of $K_{V}$ is known to be an order of $10^{21}$ mag$^{-1}$ cm$^{-2}$ at a visual wavelength 
\citep[i.e., $\lambda = 0.55 \mu$m, e.g.,][]{1978ApJ...224..132B,2010A&A...522A..84O}.
On the other hand, the effect of extinction is poorly constrained for young, accreting giant planets;
observations of these planets are currently very rare,
and hence the emitting environment remains to be studied.
Theoretically, extinction originating from gas is expected to be small at least at H$\alpha$ \citep[e.g.,][]{2022A&A...657A..38M}.
However, dust opacity can be non-negligible at optical and IR wavelengths \citep[e.g.,][]{2020MNRAS.492.3440S}.
In this work, therefore, we attempt to compute the value of $K_{\lambda}$, 
using the H$\alpha$ observations done for PDS 70 b/c and our theoretical models.

\begin{table*}
\begin{minipage}{17cm}
\begin{center}
\caption{Extinction coefficients derived from PDS 70 b/c}
\label{table3}
{\tiny
\begin{tabular}{c||cc||ccc|ccc}
\hline
                       &  \multicolumn{2}{c||}{Input parameters}                                      & \multicolumn{6}{c}{Computed quantities}                                                         \\ \hline 
                       &                      &                                                                                   & \multicolumn{3}{c|}{Accretion shock}                                                                                                                          & \multicolumn{3}{c}{Accretion flow}  \\ \cline{4-9}
                       &  Position              & Observed Line flux                                      &  Line flux                                                           &  Extinction                             & Coefficient                                   & Line flux                                                          &  Extinction                           & Coefficient                                      \\  
                      &  $r_{\rm p}$ (au)   & $L_{{\rm H}\alpha}$ (erg s$^{-1}$ cm$^{-2}$)    &  $L_{{\rm H}\alpha}$ (erg s$^{-1}$ cm$^{-2}$) &  $A_{{\rm H}\alpha}$ (mag)  & $K_{{\rm H}\alpha}$ (mag$^{-1}$ cm$^{-2}$)   & $L_{{\rm H}\alpha}$ (erg s$^{-1}$ cm$^{-2}$) & $A_{{\rm H}\alpha}$ (mag) & $K_{{\rm H}\alpha}$ (mag$^{-1}$ cm$^{-2}$)               \\  \hline
PDS 70b        & 20.2                      &  $8.3 \times 10^{-16}$                                        &  $1.0 \times 10^{-14}$                                       & 2.7                                        & $2.2 \times 10^{22}$                    & $ 5.6 \times 10^{-14}$                                     & 4.6                                       & $1.3 \times 10^{22}$                                            \\
PDS 70c        & 25.5                      &  $3.1 \times 10^{-16}$                                        &  $1.1 \times 10^{-14}$                                       & 3.8                                         & $1.6 \times 10^{22}$                   &  $5.9 \times 10^{-14}$                                    & 5.7                                        & $1.1 \times 10^{22}$             \\ 
\hline                                 
\end{tabular}
}
\end{center}
\end{minipage}
\end{table*}

Table \ref{table3} summarizes the input parameters and computed quantities.
We use equations (\ref{eq:Lacc}) and (\ref{eq:L_line}) to compute (theoretically predicted) intrinsic line flux.
The values of extinction and the coefficient are then calculated from equations (\ref{eq:ext_A1}) and (\ref{eq:ext_A2}), respectively.
The (observed) input parameters are taken from \citet{2020AJ....159..222H}. 
In addition, the stellar mass, the mass of planets b and c, and the surface density of the circumstellar disk around planet positions are assumed to be
$M_{\rm s}=0.85 M_{\odot}$, $M_{\rm p}\sim 2 M_{\rm J}$, and $\Sigma_{\rm d}^{\rm CSD} \sim 0.1$ g cm$^{-2}$, respectively, following \citet{2019A&A...625A.118K}.
The last quantity is used to compute $N_{\rm H}(=\Sigma_{\rm d}^{\rm CSD}/ m_{\rm H})$.
The distance of PDS 70 from Earth is set at 113 pc \citep{2020AJ....159..222H}. 

Our calculations show that even when the wavelength dependence of $A_{\lambda} (\propto \lambda^{-1.75})$ is taken into account \citep[e.g.,][]{2011piim.book.....D}
the coefficient $K_{{\rm H}\alpha}$ is about a few times higher than the value obtained at star-forming environments.
This is likely to be reasonable as gas contributing to extinction for accreting giants may come from the surface layer of parental circumstellar disks;
such disk gas may be poor in the dust abundance due to dust settling and growth, compared with the ISM gas.
The value of extinction itself can nonetheless be higher than that of star-forming environments simply because $N_{\rm H}$ may be much higher in planet-forming environments.
It should be pointed out that our estimate of $A_{\lambda}$ for the accretion shock case is comparable to that of \citet{2020AJ....159..222H},
where the extinction values are derived from the line flux ratio between the observed H$\alpha$ and non-detected H$\beta$.

In the following section, we will use the computed value of $K_{{\rm H}\alpha}$ to take into account extinction for the HD~163296 system.

\subsection{Comparison with theoretical prediction} \label{sec:comp}

We finally compare our theoretical predictions made in Section \ref{sec:mod} with our observational results done toward the HD~163296 system.

\begin{figure}
\begin{center}
\includegraphics[width=8.3cm]{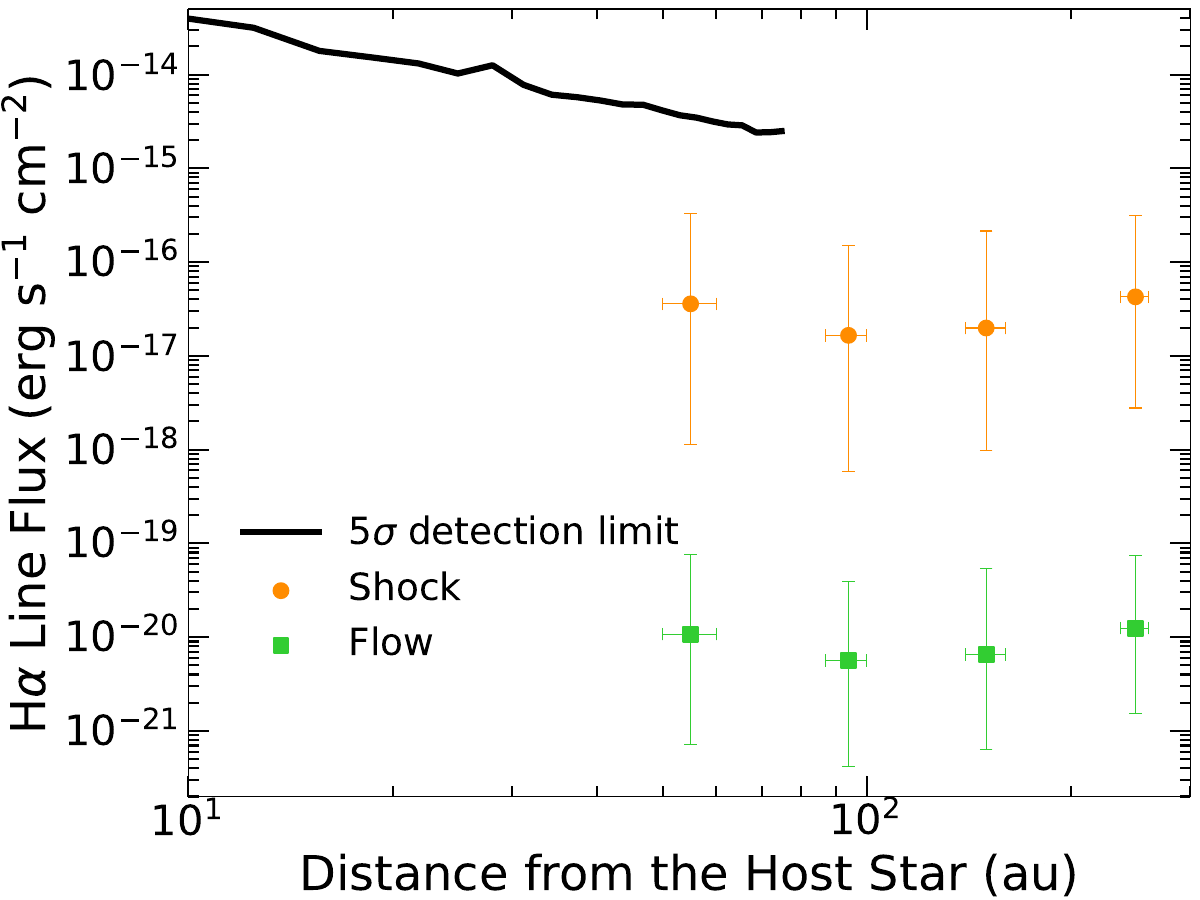}
\caption{Comparison of the theoretical predictions with the observational results.
H$\alpha$ emission from accretion shock (the orange circles) results in higher observable flux than that from accretion flow (the green squares)
due to the adopted value of $K_{{\rm H}\alpha}$ (Table \ref{table3}).
High extinction prevents careful examination of whether our theoretical predictions can reproduce the observations,
while they are not inconsistent with each other.
The current observational sensitivity needs to be improved by at least a factor of ten to reliably investigate the emission mechanisms of accreting giant planets.}
\label{fig8}
\end{center}
\end{figure}

Figure \ref{fig8} shows the results.
The line flux of H$\alpha$ for planet candidates is computed, using equations (\ref{eq:Lacc}), (\ref{eq:L_line}), (\ref{eq:ext_A1}), and (\ref{eq:ext_A2}).
The properties of these candidates are summarized in Table \ref{table1}.
The value of $\Sigma_{\rm d}^{\rm CSD} \sim 0.5$ g cm$^{-2}$ is used, following \citet{2016PhRvL.117y1101I}.
Error bars come from the ranges of planet mass and positions and the variation of $K_{{\rm H}\alpha}$ (see Table \ref{table3}).
Our calculations show that the observational results are not sensitive enough to reliably examine the theoretical predictions developed in Section \ref{sec:mod};
such an examination requires that observational sensitivity should be increased by one order of magnitude or more.
We also find that accretion shock leads to higher observed flux than accretion flow, which is expected from the value of $K_{{\rm H}\alpha}$ (Table \ref{table3}).
In addition, we have confirmed that the emission resides in the region where $T_{\rm p,e} \simeq T_{\rm therm}$ (see Figure \ref{fig5}),
and hence if H$\alpha$ would be observed toward the HD~163296 system,
then the line could be used as a direct probe of the disk-limited gas accretion stage of giant planet formation.

It can thus be concluded that hydrogen emission lines, especially H$\alpha$, are useful tracers of whether giant planets undergo magnetospheric accretion at their final formation stages.
However, the current observational capability may not be high enough to reliably test the emission mechanisms (e.g., accretion shock vs accretion flow);
when planets are embedded in actively accreting circumstellar disks, the emission from planets itself can be strong.
Since the emission is the direct outcome of high accretion flow onto planets, the flow in turn attenuates the observed flux significantly.
When planets are in disks with low stellar accretion rates, the emission becomes weaker, which simply makes it difficult to be observed.
Improvement of observational sensitivity by a factor of ten or more will open up a promising window to carefully investigate the final giant planet formation stage.

\subsection{Implications for other lines}

As described above, H$\alpha$ lines tend to suffer from extinction significantly.
We here explore other lines (e.g., Pa$\beta$ and Br$\gamma$),
which are less attenuated by magnetospheric accretion flow.

\begin{figure*}
\begin{minipage}{17cm}
\begin{center}
\includegraphics[width=8.3cm]{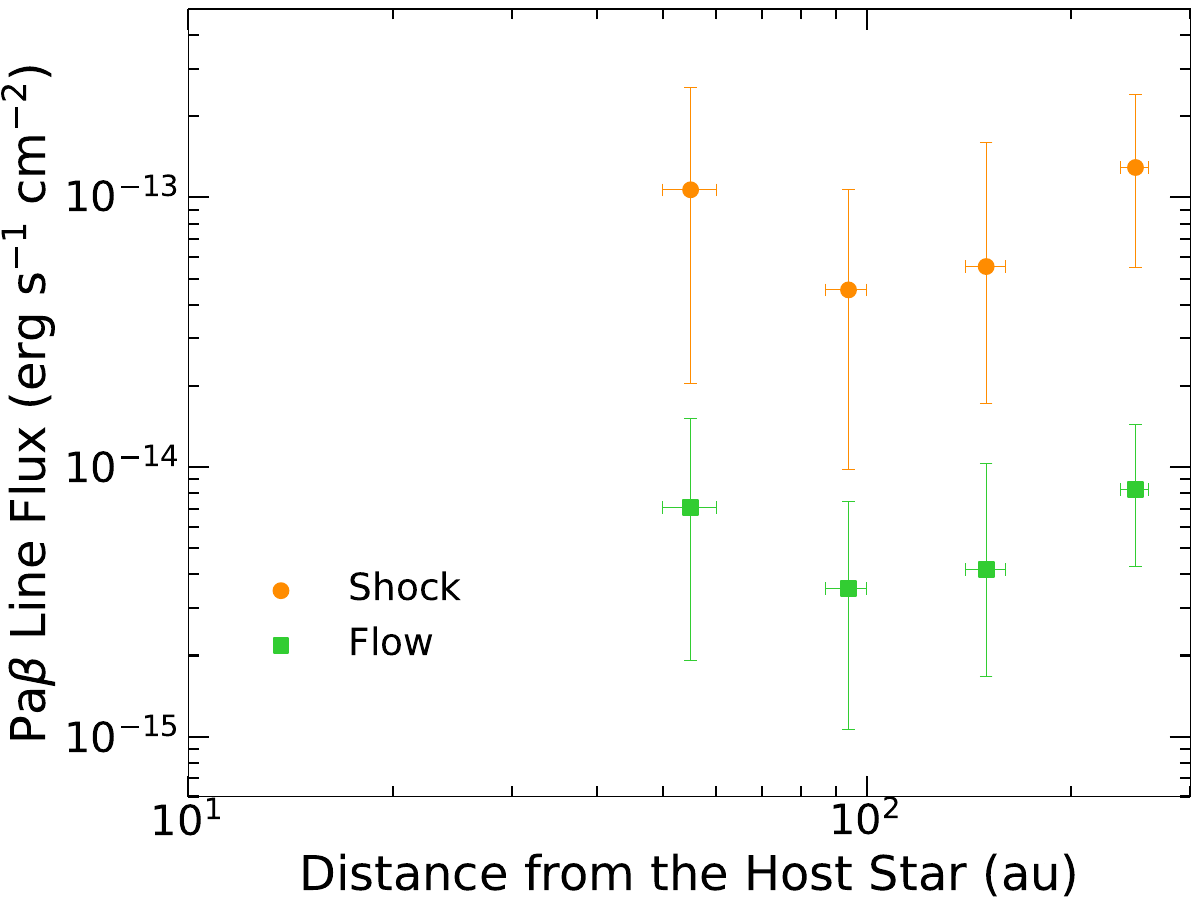}
\includegraphics[width=8.3cm]{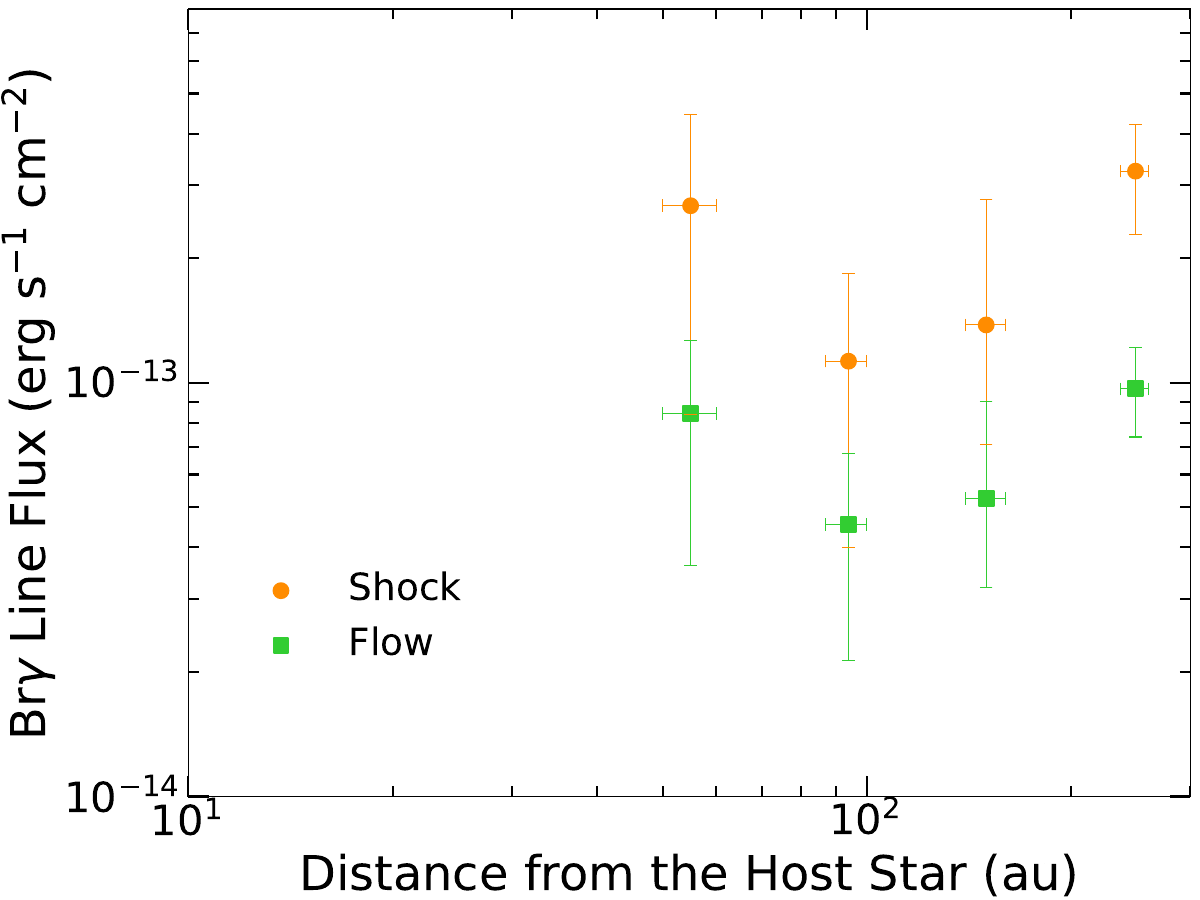}
\caption{Predicted line flux for Pa$\beta$ and Br$\gamma$ on the left and right panels, respectively, as done in Figure \ref{fig8}.
Hydrogen lines at longer wavelengths tend to be observed more readily as the effect of extinction becomes weaker.
Multi-band observations are crucial not only for discovering accreting giants, but also for characterizing them.}
\label{fig9}
\end{center}
\end{minipage}
\end{figure*}

Figure \ref{fig9} shows the resulting line flux for Pa$\beta$ and Br$\gamma$.
In these calculations, we adopt the same input parameters as done in Section \ref{sec:comp}.
We find that the observable line flux for Pa$\beta$ and Br$\gamma$ should be much higher than that of H$\alpha$.
This arises simply because extinction is a decreasing function of $\lambda$ (i.e., $A_{\lambda} \propto \lambda^{-1.75}$).
We have confirmed that contamination from continuum emission by planets and the disk is negligible for the HD~163296 system.

Thus, intrinsic hydrogen emission lines originating from accreting giants are weaker with increasing wavelengths (see Figure \ref{fig5}).
However, the effect of extinction also becomes weaker for longer wavelengths.
As a result, the observability of these lines (e.g., Pa$\beta$ and Br$\gamma$) becomes higher than that of H$\alpha$.
Multi-band observations will expand the possibility of discovering and characterizing young giant planets embedded in parental circumstellar disks,
which may undergo magnetospheric accretion.

\section{Discussion} \label{sec:disc}

Our theoretical model has been developed, based on physical arguments and existing studies in the literature.
However, it is a very simple model, and more investigations are required to verify our predictions.
Here, we summarize key assumptions adopted in this work and potential caveats relevant to the assumptions.

First, we discuss the effective temperature of accreting planets.
In the above sections, we have considered two limiting cases: $T_{\rm p,e} \simeq T_{\rm int}$ and $T_{\rm p,e} \simeq T_{\rm therm}$.
This essentially assumes that planetary magnetic fields and the resulting magnetospheric accretion are regulated purely by one of the temperatures.
In reality, both temperatures could affect them.
One can estimate this effect by re-writing equation (\ref{eq:T_pe}) as
\begin{eqnarray}
\label{eq:T_pe1}
T_{\rm p,e}  & =  & \left( T_{\rm int}^4 +  T_{\rm therm}^4 \right)^{1/4} = T_{\rm therm}  \left[1 + \left( \frac{ T_{\rm int}  }{ T_{\rm therm} } \right)^4 \right]^{1/4} \\ \nonumber
                    & \equiv & f_{\rm cor} T_{\rm therm} ,
\end{eqnarray}
where $f_{\rm cor} = [ 1 + ( T_{\rm int} /T_{\rm therm})^4]^{1/4}$ is the correction factor.
Without loss of generality, one can focus on the case that $ T_{\rm int} \le T_{\rm therm} $.
Then the factor takes a maximum value when $ T_{\rm int} =T_{\rm therm} $, leading to  $f_{\rm cor} \simeq 1.19$.
The resulting difference in the strength of planetary magnetic fields is $1.26$ (see equation (\ref{eq:B_ps})).
This $\la 30$ \% difference would not be significant for this work
as our model is very simple and the values of physical parameters are not constrained tightly.
We therefore conclude that considering two limiting cases would be useful
and even if the other contribution would be taken into account,
our results would not change very much.
It should be noted that as discussed in \ref{sec:line_lum}, 
reliable differentiation of the two limiting cases is possible 
only when stellar accretion rates are high or low (the dashed and solid lines in Figure \ref{fig4});
in between, both two cases are possible, 
and our model cannot reliably determine which temperature ($ T_{\rm int}$ vs $T_{\rm therm} $) would play a dominant role in regulating magnetospheric accretion.

Second, we discuss the feasibility of magnetospheric accretion for accreting planets.
We have so far assumed that magnetospheric accretion is realized 
if planetary magnetic fields are sufficiently strong.
This is the very minimum requirement, however.
In fact, the disk gas in the vicinity of planets needs to be ionized enough,
so that the disk gas can be well coupled with planetary magnetic fields.
Such a condition can be met for PDS 70 b/c \citep{2021ApJ...923...27H},
and hence it would be possible for other accreting giant planets.
However, it is not obvious.
Explicit confirmation is desired for the HD 163296 system.

Third, this work targets giant planets embedded in circumstellar disks and 
assumes steady state accretion from circumstallar disks to circumplanetary disks and down to planets.
It is possible that giant planets surrounded by circumplanetary disks are isolated from their parental circumstellar disks.
In fact, some observations discover such targets \citep[e.g., GQ Lup and Delorme 1b;][]{2021AJ....162..286S,2022ApJ...935L..18B,2023A&A...669L..12R}.
A more comprehensive list of accreting substellar objects, including companions, as well as their accretion rates is available at \citet{2023AJ....166..262B}.
If giant planets are isolated from circumstellar disks, 
circumplanetary disks are not replenished by circumstellar disks, and accretion rates onto planets and stars are not correlated with each other.
Our model cannot be applied to such systems.
Also, even if giant planets and their circumplanetary disks are embedded in the circumstellar disks,
it is not guaranteed that the steady state accretion assumption would hold for them.
If the systems would undergo episodic accretion, then our model provides only an intermediate value for the accretion luminosity and line flux.

Finally, we discuss extinction.
As pointed out in Section \ref{sec:extinction},
extinction is one of the most unexplored areas in the literature.
We have obtained the value from PDS 70 b/c and applied it to the HD 163296 system.
This involves with two implicit assumptions.
The first assumption is that the extinction value derived only from H$\alpha$ observations is reasonable at other wavelenghts,
and the other one is that extinction would be comparable for both the PDS 70 and HD 163296 systems.
We here examine the validity of these two assumptions.

The first assumption can be verified by comparing other observations.
For instance, \citet{2021AJ....162..214U} conducted Keck/OSIRIS observations to search for Pa$\beta$ emission lines from the PDS 70 system.
They did not detect any emission and derived the 5$\sigma$ detection limits as done in this work,
which are $1.4 \times 10^{-16}$ erg s$^{-1}$ cm$^{-2}$ and $1.9 \times 10^{-16}$ erg s$^{-1}$ cm$^{-2}$ for PDS 70 b and c, respectively.
Adopting the extinction value derived from H$\alpha$ emission (Table \ref{table3}),
our model predicts Pa$\beta$ emission line fluxes to be $2.5 \times 10^{-16}$ erg s$^{-1}$ cm$^{-2}$ and $2.0 \times 10^{-16}$ erg s$^{-1}$ cm$^{-2}$ for PDS 70 b and c, respectively
in the accretion shock case.
On the other hand, predicted Pa$\beta$ emission line fluxes become $8.1 \times 10^{-16}$ erg s$^{-1}$ cm$^{-2}$ and $6.2 \times 10^{-16}$ erg s$^{-1}$ cm$^{-2}$ for PDS 70 b and c, respectively
in the accretion flow case.
According to the face values, our model implies that accretion shock would be a most likely scenario.
However, given caveats discussed above and uncertainties in physical parameters,
more detailed investigations are required to derive solid determination.
Instead, since our flux estimates derived from the simple model are comparable to the observationally inferred limits,
it might not be unrealistic to consider that the extinction value derived only from H$\alpha$ emission would be reasonable at other wavelengths as well.

It should be pointed out that our extinction values are much higher than the values known for these systems, 
which are $A_V$ of $\sim 0.05$ for the PDS 70 system \citep{2018A&A...617L...2M} 
and $A_V$ of $\la 0.5$ for the HD 163296 system \citep[e.g.,][]{2019ApJ...875...38R}.
This thus suggests that giant planets embedded in their circumstellar disks tend to be further obscured by surrounding planet-forming materials.

For the second assumption, we must admit that it cannot be examined readily currently.
This is mainly because of the lack of observations as discussed above.
The extinction value adopted in this work should be viewed as a reference,
and the resulting line fluxes could change significantly.
Our model can provide better flux estimates once extinction at planet-forming environments is constrained tightly,
and/or can be used to constrain extinction itself if multi-band observations and the resulting line fluxes are available.

\section{Summary \& Conclusions} \label{sec:sum}

We have investigated theoretically when accreting giant planets embedded in circumstellar disks emit observable hydrogen lines via magnetospheric accretion.
Our theoretical predictions have been compared with our observations that are conducted for HD 163296.
This target star hosts the circumstellar disk exhibiting the gas and dust gap structures as well as meridional flows.
These disk structures are widely considered as potential signatures of ongoing giant planet formation.
Our efforts have been made, in order to increase the sample size of confirmed, young giant planets and to quantify the ubiquity of H$\alpha$ emission from these planets.

We have begun our exploration from developing a theoretical model (Section \ref{sec:mod}).
We have first examined energetics of accreting giant planets (Figure \ref{fig1}); 
some of accretion energy affects the effective temperature of these planets (equation (\ref{eq:T_pe})).
By using a simple scaling law, magnetic fields of accreting giants have been computed (equation (\ref{eq:B_ps})).
We have found that depending on how the effective temperature of planets is determined,
two cases can be considered separately;
when the effective temperature is regulated mainly by earlier formation histories,
the temperature becomes the fundamental parameter of whether magnetospheric accretion occurs (equation (\ref{eq:B_ps1})).
On the other hand, when disk-limited gas accretion becomes energetic enough to affect the effective temperature,
magnetospheric accretion and the accompanying hydrogen line emission can be self-regulating (equation (\ref{eq:B_ps2})).

We have then examined under what conditions, magnetospheric accretion is realized.
Under the assumption that the magnetic pressure of planets is balanced with the ram pressure of accreting disk gas,
we have computed all the key quantities such as the effective temperature, magnetic field, accretion rate, and accretion luminosity of planets (Figure \ref{fig2}).
The resulting values are expressed as a function of both the planet mass and the location of the inner edge of truncated disks.
We have also constrained the location of disks' inner edge, 
by considering the conservation of energy for magnetospheric accretion as well as a global configuration of accretion flow.
If giant planets achieve a steady state, which is suggested for PDS 70 b/c,
then the condition that magnetospheric accretion becomes possible is derived as a function of the planet mass and stellar accretion rates (Figure \ref{fig3}).
This condition divides the parameter space of the planet mass and stellar accretion rate into three regions (Figure \ref{fig4}):
when stellar accretion rates are sufficiently high (equation (\ref{eq:Mdot_s_int})), 
magnetospheric accretion controls the corresponding accretion heating,
and the resulting accretion luminosity is the outcome of such a self-regulating process; and
when stellar accretion rates are low and planets are massive (equation (\ref{eq:Mdot_s_therm})),
earlier formation histories determine whether magnetospheric accretion occurs.
There is an intermediate region in which both cases are possible.

We have computed hydrogen line luminosities, using relationships between the accretion and line luminosities.
Two relationships have been adopted in this work (Table \ref{table2}): 
one is derived from theoretical studies,
where hydrogen lines are produced at planetary surfaces due to accretion shock;
and the other is based on observations of young stellar objects,
where hydrogen lines come from magnetospheric accretion flow.
These relationships lead to higher line luminosities from accretion flow than those from accretion shock (Figure \ref{fig5}).
Also, line luminosities decrease with increasing wavelengths (i.e., from H$\alpha$ to Pa$\beta$ and up to Br$\gamma$).

We have conducted new observations targeting HD 163296 with Subaru/SCExAO+VAMPIRES (Section \ref{sec:data}).
Our observations did not detect any point-like source emitting H$\alpha$ (Figure \ref{fig6}).
In order to compare our theoretically computed H$\alpha$ line flux with the observations,
we have estimated the 5$\sigma$ detection limit (Figure \ref{fig7}).
Also, we have quantified the effect of extinction, 
by applying our theoretical model to the observed H$\alpha$ emission of PDS 70 b/c (Table \ref{table3}).

We have finally compared our theoretical results with observational ones,
and found that our observations are not sensitive enough to reliably examine our theoretical predictions (Figure \ref{fig8}).
Our theoretical model has been applied to giant planet candidates, 
which are suggested from various observational signatures (Table \ref{table1}).
Reliable verification of our theoretical predictions can be done
if observational sensitivity will be improved by a factor of ten or more. 
We have also computed the line flux of Pa$\beta$ and Br$\gamma$
and shown that the observable flux increases with increasing wavelengths (Figure \ref{fig9}).
This is the direct outcome of extinction.
Inclusion of extinction leads to higher line flux from accretion shock than that from accretion flow,
which is opposite to the theoretical prediction without extinction. 

We have focused on magnetospheric accretion as a plausible mechanism of emitting hydrogen lines from accreting giant planets.
In the literature, other mechanisms have been proposed.
For instance, H$\alpha$ emission may be generated from the surface layer of either planets or circumplanetary disks without truncating the disks.
This becomes possible,
if the infall gas from circumstellar disks hit their surface layers directly \citep[e.g.,][]{2018ApJ...866...84A,2020ApJ...902..126S,2021ApJ...921...10T}.
More detailed models are required to comprehensively explore the hydrogen emission mechanism of accreting giant planets embedded in their circumstellar disks.

In conclusion, hydrogen emission lines can be a useful probe of the final stage of giant planet formation.
However, H$\alpha$ tends to suffer from extinction considerably, especially for giant planets deeply embedded in their parental circumstellar disks.
Multi-band observations (e.g., Pa$\beta$ and Br$\gamma$) are necessary 
to efficiently discover young, accreting giant planets and carefully examine the origin of hydrogen emission lines from these planets.
Ongoing and planned JWST observations can play a leading role on this topic \citep[e.g.,][]{2023ApJ...949L..36L}.

\begin{acknowledgments}

The authors thank an anonymous referee for useful comments on our manuscript.
This research was carried out in part at the Jet Propulsion Laboratory, California Institute of Technology, 
under a contract with the National Aeronautics and Space Administration (80NM0018D0004),
and funded by a Keck Principal Investigator Data Award (KPDA), managed by NExScI for NASA.
NASA Keck time is administered by the NASA Exoplanet Science Institute. 
Data presented herein were obtained at the W. M. Keck Observatory from telescope time allocated to the National Aeronautics and Space Administration 
through the agency's scientific partnership with the California Institute of Technology and the University of California. 
The Observatory was made possible by the generous financial support of the W. M. Keck Foundation.
The authors wish to recognize and acknowledge the very significant cultural role and reverence that 
the summit of Mauna Kea has always had within the indigenous Hawaiian community. 
We are most fortunate to have the opportunity to conduct observations from this mountain.
Y.H. is supported by JPL/Caltech.
T.U. is partially supported by Grant-in-Aid for Japan Society for the Promotion of Science (JSPS) Fellow, JSPS KAKENHI Grant No. JP21J01220, 
and NASA ROSES XRP award 80NSSC19K029.
M.T. is supported by JSPS KAKENHI Grant Nos. 18H05442, 15H02063, and 22000005. 

\end{acknowledgments}

\bibliographystyle{aasjournal}
\bibliography{adsbibliography}    



\end{document}